\documentclass[%
 aapm,
 mph,%
 amsmath,amssymb,
 reprint,%
 superscriptaddress%,
]{revtex4-2}

\usepackage{graphicx}% Include figure files
\usepackage{dcolumn}% Align table columns on decimal point
\usepackage{color} % for the command \textcolor
\usepackage{bm}% bold math
\usepackage{soul} % for the command \hl

\usepackage[mathlines]{lineno}% Enable numbering of text and display math

\begin{document}

% \title[PRISM 2.0]{An Even Faster Algorithm for Scanning Transmission Electron Microscopy (STEM) Imaging and Diffraction Simulations}

%Maria1: specify electron? 
\title{Random Forest Prediction of Crystal Structure from Electron Diffraction Patterns Incorporating Multiple Scattering}

\author{Samuel P. Gleason}
%\email{smglsn12@berkeley.edu}
\affiliation{National Center for Electron Microscopy, Molecular Foundry, Lawrence Berkeley National Laboratory, 1 Cyclotron Road, Berkeley, CA, USA, 94720}
\affiliation{Department of Chemistry, University of California, Berkeley, CA, USA}

\author{Alexander Rakowski}
% \email{ARakowski@lbl.gov }
\affiliation{National Center for Electron Microscopy, Molecular Foundry, Lawrence Berkeley National Laboratory, 1 Cyclotron Road, Berkeley, CA, USA, 94720}

\author{Stephanie M. Ribet}
\affiliation{National Center for Electron Microscopy, Molecular Foundry, Lawrence Berkeley National Laboratory, 1 Cyclotron Road, Berkeley, CA, USA, 94720}

\author{Steven E. Zeltmann}
% \email{}
\affiliation{School of Applied and Engineering Physics, Cornell University, Ithaca, NY, 14853, USA}
\affiliation{PARADIM, Materials Science \& Engineering Department, Cornell University, Ithaca, NY, 14853, USA}

\author{Benjamin H. Savitzky}
% \email{}
\affiliation{National Center for Electron Microscopy, Molecular Foundry, Lawrence Berkeley National Laboratory, 1 Cyclotron Road, Berkeley, CA, USA, 94720}

\author{Matthew Henderson}
% \email{}
\affiliation{National Center for Electron Microscopy, Molecular Foundry, Lawrence Berkeley National Laboratory, 1 Cyclotron Road, Berkeley, CA, USA, 94720}

\author{Jim Ciston}
% \email{jciston@lbl.gov}
\affiliation{National Center for Electron Microscopy, Molecular Foundry, Lawrence Berkeley National Laboratory, 1 Cyclotron Road, Berkeley, CA, USA, 94720}

\author{Colin Ophus}
%\email{clophus@lbl.gov}
\affiliation{National Center for Electron Microscopy, Molecular Foundry, Lawrence Berkeley National Laboratory, 1 Cyclotron Road, Berkeley, CA, USA, 94720}

\date{\today}

\begin{abstract}

Diffraction is the most common method to solve for unknown or partially known crystal structures. However, it remains a challenge to determine the crystal structure of a new material that may have nanoscale size or heterogeneities. Here we train an architecture of hierarchical random forest models capable of predicting the crystal system, space group, and lattice parameters from one or more unknown 2D electron diffraction patterns. Our initial model correctly identifies the crystal system of a simulated electron diffraction pattern from a 20 nm thick specimen of arbitrary orientation 67\% of the time. We achieve a topline accuracy of 79\% when aggregating predictions from 10 patterns of the same material but different zone axes. The space group and lattice predictions range from 70-90\% accuracy and median errors of 0.01-0.5$\AA{}$, respectively, for cubic, hexagonal, trigonal and tetragonal crystal systems while being less reliable on orthorhombic and monoclinic systems. We apply this architecture to a 4D-STEM scan of gold nanoparticles, where it accurately predicts the crystal structure and lattice constants. These random forest models can be used to significantly accelerate the analysis of electron diffraction patterns, particularly in the case of unknown crystal structures. Additionally, due to the speed of inference, these models could be integrated into live TEM experiments, allowing real-time labeling of a specimen.

\vspace{\baselineskip}

\end{abstract}
% \pacs{PACS Numbers}
% \keywords{Keywords go here}
\maketitle

\section*{Introduction}

Many technologically important materials are crystalline, formed from small unit cells tiled periodically in three dimensions along a lattice \citep{sands1993introduction}. 
The arrangement of atoms inside the unit cell and the shape of the lattice control many important materials properties \citep{desiraju2010crystal}. Examples include catalytic activity \citep{shen2021crystal}, ion diffusion kinetics \citep{ji2019hidden}, optical properties \citep{bhaumik2020recent}, yield strength \citep{senkov2016development}, plasticity \citep{butler2018mechanisms} and magnetism \citep{eriksson1990orbital}.
We therefore require robust characterization methods to predict \citep{curtarolo2003predicting}, experimentally determine \citep{harris1996crystal} and refine crystal structures \citep{palatinus2015structure, griesemer2021high}.

%crystal structure controls many important material science properties; for example 

Transmission electron microscopy (TEM) and scanning TEM (STEM) provide several methods which can be used to determine crystal structures \citep{ophus2023quantitative}. These range from direct imaging at atomic resolution in 2D \citep{Meyer2007_the_structure} and 3D \citep{yang2017deciphering}, spectroscopy in 2D \citep{lu2014atomic, wenner2017atomic} and 3D \citep{jenkinson2022multimode}, and diffraction in 2D \citep{martynowycz2018electron} and 3D \citep{nannenga2019cryo}.
Electron diffraction is the most common electron microscopy (EM) method to solve for unknown or partially known crystal structures \citep{zuo2013electron}, as it provides direct and interpretable feedback on the orientation of crystal grains under the beam. Electron diffraction is often invaluable when high spatial resolution is required, due the sub angstrom spatial resolution possible with STEM scans. The ability to utilize STEM and collect a diffraction pattern at every probe position, with a spatial resolution orders of magnitude higher than possible with even the most focused X-ray beams, is one of the key advantages of using electron diffraction for crystal structure determination \citep{ophus2023quantitative, Gemmi2019Electron}. Coherent scattering from a single crystal aligned along a zone axis will produce a two-dimensional Bragg diffraction pattern, which is composed of sharp spots positioned at lattice crossings. 
The spot positions are defined by the intersection of the Ewald sphere with the reciprocal lattice points of the crystal, which are given by the three-dimensional Fourier transform of the real space lattice  \citep{authier2006dynamical}. 
The diffraction spot intensities, integrated over the beam shape, are defined by the shape function of the parent crystal evaluated at the closest distance, known as the excitation error, of the Ewald sphere for each reciprocal lattice point \citep{deGraef2003introduction}.

Most existing analytical and simulation techniques to analyze electron diffraction patterns make use of the kinematical approximation, which assumes the scattering particles scatter only once within the crystal \citep{deGraef2003introduction, Gruene2021Establishing, ophus2022automated}. This results in diffraction spot intensities that vary linearly with the structure factor intensities \citep{zachariasen1967general}. By contrast, spot intensities in electron diffraction patterns tend to be strongly modulated by multiple scattering of the beam, where electrons scatter off of the crystal multiple times during transmission. The higher incidence of multiple scattering in electron diffraction is mainly due to the far stronger interaction between electrons and the crystalline material vs X-rays \citep{Eggeman2012Advances}. This causes non-trivial changes in diffraction spot intensities, and these deviations from kinematical theory are unique to each specimen thickness and crystal orientation to the electron beam \citep{Zeltmann2023blochwave, Gruene2021Establishing}. Multiple scattering can also result in spots at locations of forbidden reflections using kinematical theory, significantly complicating the application of  kinematical theory to electron diffraction analysis \citep{Klar2023Accurate, Mahr2022Towards, Gruene2021Establishing}.  
 
\par
While there are programs that attempt to solve for structural information from dynamical electron diffraction patterns, these brute force simulations are very computationally intensive and require pre-existing structural knowledge about the potential systems present \citep{palatinus2015structure}. These challenges can be somewhat mitigated by utilizing precession electron diffraction (PED), where the incident electron beam is rotated around the central axis of the microscope and a diffraction pattern taken at each rotation angle. This results in PED pattern, which is a summation of each of these individual patterns, and therefore serves to limit the contributions of dynamical scattering effects to the full PED pattern and renders kinematical theory more suitable \citep{Gemmi2019Electron, Jeong2021Automated, Eggeman2012Advances, Oleynikov2007Precession}. However, since collecting PED patterns requires specific hardware that many TEMs do not possess, dynamical diffraction patterns are still quite common and remain much more difficult to analyze. Due to the deviations between dynamical and kinematical diffraction, models trained using electron diffraction patterns simulated using the kinematical approximation will struggle significantly to predict patterns containing reflections generated by multiple scattering events. To mitigate this issue, we have generated a dataset of simulated dynamical diffraction patterns, with a wide range of thicknesses and crystal orientations, to ensure our model is prepared for a broad of a set of real world experimental conditions.

Due to the widespread adoption of fast direct electron detectors, we can now quickly measure thousands or even millions of diffraction patterns from a sample \citep{maclaren2020comparison, Ciston20194DCamera}. These datasets are typically comprised of 2D diffraction patterns recorded over a 2D grid of converged electron probe positions, producing a four-dimensional (4D)-STEM dataset \citep{ophus2019four}. The simplest method to determine the sample's crystal structure from these diffraction patterns is to generate libraries of diffraction patterns at different orientations from one or more potential crystal structures \citep{zaefferer1994line}. 
These simulated diffraction patterns are then matched against experiments to determine both the best fit orientation and phase \citep{rauch2005rapid}.
There are several algorithms and software packages which can perform this automated crystal orientation matching (ACOM) \citep{kobler2017challenges, ophus2022automated, cautaerts2022free}. 
However, this method cannot be used to determine crystal structure or orientation from structures not included in the diffraction libraries.
It is possible to jointly solve for the relative orientations of polycrystalline grains and use 3D electron diffraction methods to solve for the structure \citep{gallagher2020atomic}, but this requires a sufficiently large number of orientations and extended analysis \citep{nannenga2014protein}.
 
Large numbers of experimental diffraction patterns can also be analyzed with machine learning (ML) methods \citep{kalinin2022machine}. Unsupervised ML has been used for clustering large volumes of electron diffraction patterns and unmixing patterns containing multiple grains \citep{martineau2019unsupervised, Bruefach2023Robust, matinyan2023machine}. These approaches are valuable for determining regions of similarity in a sample and for simplifying analysis of diffraction patterns which are superpositions of multiple grains, as these can be challenging to analyze without further processing. However, clustering methods are unable to perform tasks such as assigning a crystal system or lattice parameter values to an unlabeled electron diffraction pattern. Additionally, variational autoencoders have been used to develop a latent representation for electron diffraction patterns which can be used to cluster similar patterns \citep{Oxley2021Probing, Bridger2023Versatile} and train subsequent ML models to make predictions about physical properties of the sample when the latent space represents convergent beam low energy electron diffraction (CBLEED) \citep{Ivanov2024Autoencoder}. However, these methods have yet to be applied broadly to direct prediction of crystal system, lattice parameters or space groups of crystals from their electron diffraction patterns.

The most common architecture for supervised ML electron diffraction analysis is the convolutional neural network (CNN), which has achieved a wide variety of applications when analyzing electron diffraction patterns. These applications include precise strain and orientation mapping from known structures \citep{yuan2021training}, inverting complex dynamical diffraction patterns into projected kinematical structure factors \citep{munshi2022disentangling}, determining crystal symmetries in electron backscattered diffraction (EBSD)
\citep{kaufmann2020crystal}, and predicting space groups when focusing specifically on cubic materials and a subset of cubic space groups \citep{Ra2021Classification}. More broad crystal system classification tasks have also been conducted with CNNs, such as hybrid models predicting crystal structure from a inputs containing a combination of 1D and 2D electron diffraction profiles \citep{Aguiar2019Decoding}. Among the most robust crystal system prediction methods is\citep{Chen2023Automated}, which generated a dataset of roughly half a million simulated kinematical electron diffraction patterns and constructed a CNN for crystal system prediction directly from a single input 2D electron diffraction pattern, attaining an accuracy of 0.55 for simulated kinematical data. They also developed a prediction aggregation procedure, which is based on continual pattern acquisition until a specified level of uniformity in the predictions is reached. This procedure attains an accuracy of 94\% on simulated kinematical data. Although this aggregated accuracy is quite high, the requirement of continual data acquisition until a near uniform set of predictions across the individually acquired patterns is reached requires that the model be integrated into the data acquisition procedure to fully utilize this aggregation procedure.

In this work we build a hierarchical architecture of random forest models to predict crystal system, lattice parameters and space group from electron diffraction patterns. We have selected this architecture mainly due to the ability of an ensemble of estimators to generate a prediction uncertainty metric. Specifically, in this work we derive a confidence metric for classification tasks based on the number of decision trees predicting the most common label, subtracted from the number of decision trees predicting the second most common label, and divided by the total number of decision trees. This provides a significant advantage over neural network architectures, which are typically black box estimators, and previous studies have shown the quantitative utility of utilizing the prediction confidence derived from random forest models \citep{Gleason2023Prediction}. This is particularly valuable in the case of predicting crystal structure from electron diffraction patterns due to the inherent uncertainty in deriving 3D information from a 2D representation of the crystal, as ambiguity in the diffraction pattern at some crystal orientations can cause misidentifications even by trained experts \citep{Ponce2021advances, Zachman2022measuring}. Knowing that there is a low vs high degree of uncertainty in the crystal structure prediction can yield valuable information to the user, particularly if multiple patterns from the same material are available. 

In this study, we simulate electron diffraction patterns from 36,600 crystals, using 100 symmetry-unique orientations and 100 specimen thicknesses from 2-200 nm for each crystal. We use the Bloch wave method to include multiple scattering from thick samples in our diffraction pattern simulations \citep{deGraef2003introduction}. We parameterize these diffraction patterns using a polar coordinate basis. Using this parameterized representation, we develop a random forest ML architecture which takes this basis as an input and returns predictions of crystal system, space group, and lattice parameter lengths. We validate our method on individual diffraction patterns and test the accuracy of our method applied to aggregate sets of diffraction patterns to mimic the inputs from a 4D-STEM scan of a polycrystalline sample. We also validate the use of this architecture as an experimental analysis tool by predicting a 4D-STEM scan of a sample of gold nanoparticles (AuNPs). Finally, we have made all of our simulated diffraction patterns freely available to researchers who want to develop their own automated crystal structure analysis methods.

\begin{figure*}[ht]
    \centering
    \includegraphics[width=12cm]{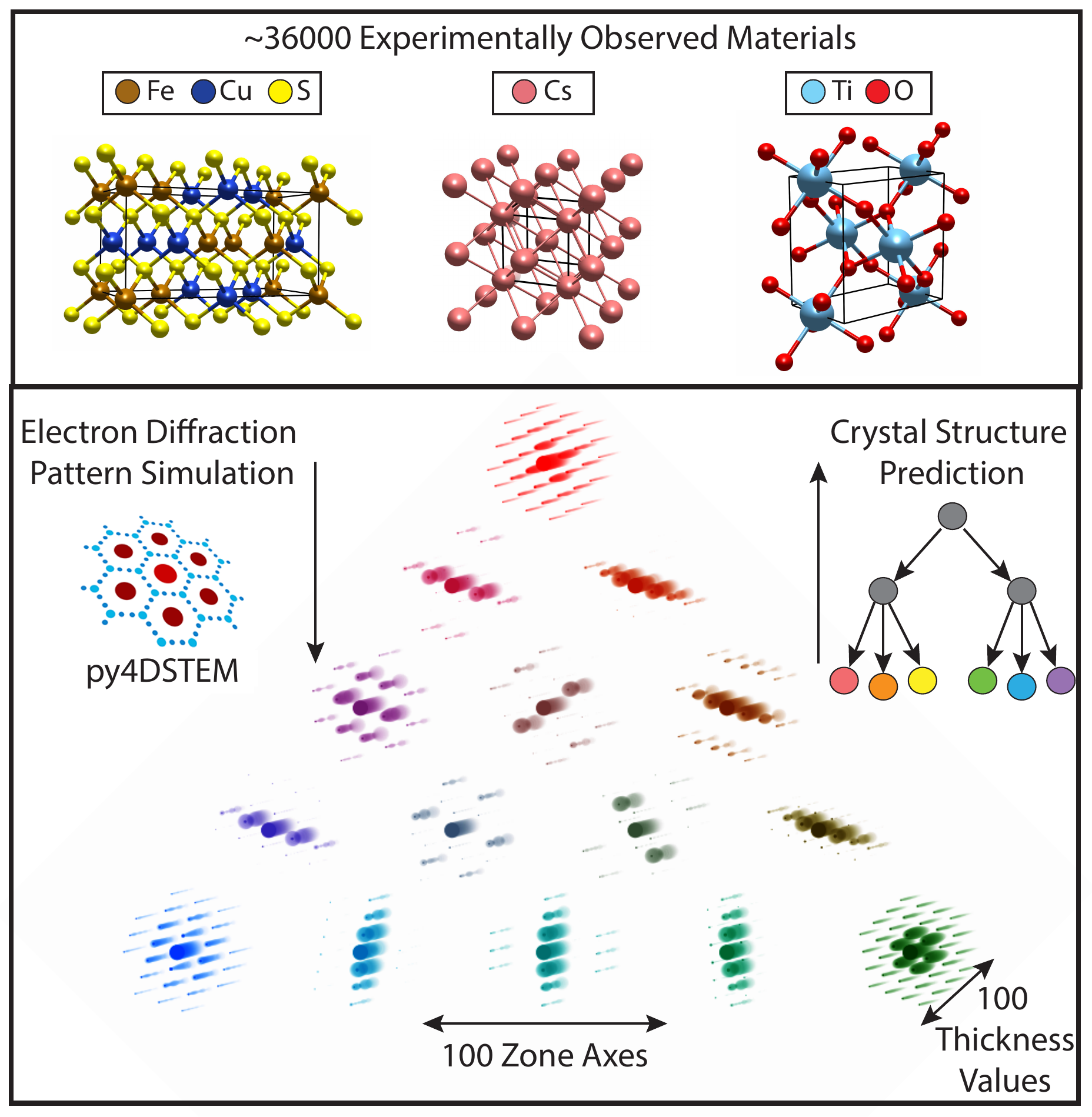}
    \caption{Summary of the generation of data  used to train this model. Top row: 36000 structures are extracted from the materials project, filtering for only materials that have been experimentally observed. Bottom row: For each of these materials, 100 unique zone axes are simulated using py4DSTEM by generating a crystal system specific orientation map to ensure the symmetrical uniqueness of each zone axis. For each unique zone axis, 100 specimen thicknesses are simulated, from 2-200 nm with 2 nm spacing. The thicknesses are indicated by the stack of patterns making up each unique zone axis.}
    \label{fig_full_scope} %This is for you to refer to the figure in maintext text
\end{figure*}

\section*{Methods}

\subsection*{Diffraction Pattern Dataset Generation}

\begin{figure*}[ht]
    \centering
    \includegraphics[width=10cm]{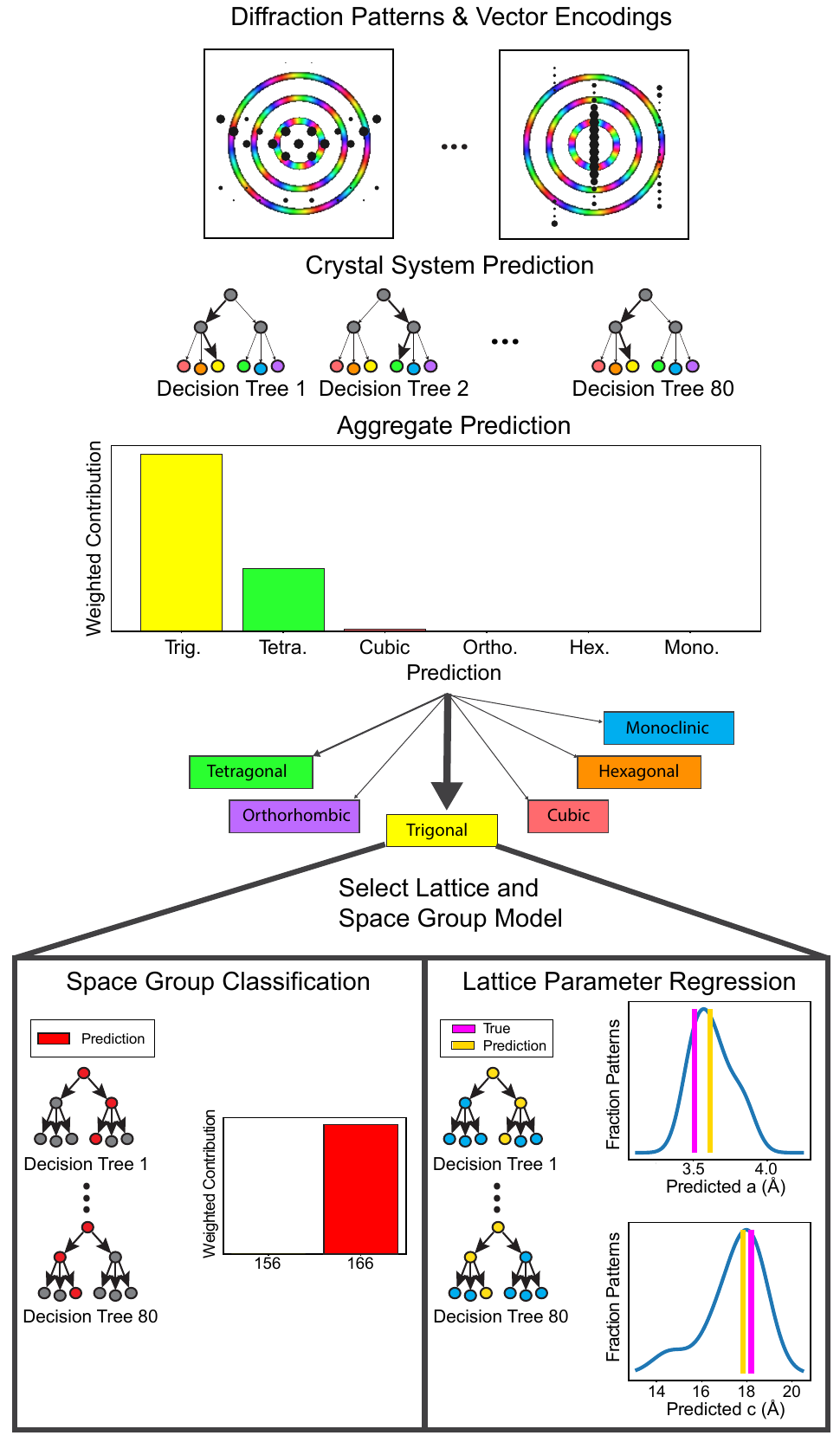}
    \caption{Illustration of the full model architecture. First, the radial basis transformation is applied to the input diffraction pattern(s), represented as Bragg lists. The black spots in the first row show the location of diffraction spots, scaled by their intensities, and the rings show example radial/angular bin pairs with 6 fold angular symmetry ranging from low (center) to high (outer) radial order. The color of each radial basis bin represents the complex phase. Then, the crystal system is predicted and, based on this prediction, the pattern(s) are transferred to the lattice and space group prediction models corresponding to the predicted crystal system.}
    \label{fig_model_outline} %This is for you to refer to the figure in maintext text
\end{figure*}

To generate our simulated electron diffraction pattern library we obtained crystal structures from the Materials Project \citep{jain2013commentary}, including only structures which have been experimentally observed. This generated a list of $\approx$50,000 materials. We then removed materials with unit cell volumes greater than 1,000 \AA{}$^3$. This was done to ensure efficient simulation and because materials with unit cells this large rarely occur in nature. This produced a dataset containing $\approx$36,000 materials. This dataset contains materials from all 7 crystal systems, and their distributions can be seen in Figure S1. It is worth noting that the crystal system distribution is not uniform, which can be considered an advantage of this model, as it carries an implicit bias towards more experimentally observed crystal structures (i.e. orthorhombic) and away from more uncommon ones (i.e. trigonal). For each crystal, 100 unique and non-symmetrically redundant zone axis were selected, spanning the crystals' fundamental zone. We note that the sampling density for high symmetry crystals (e.g. cubic) will therefore necessarily be higher than those for low symmetry crystals (e.g. monoclinic) \citep{ophus2022automated}. An overview of our dataset generation procedure and the scope of our dataset can be found in Figure \ref{fig_full_scope}.

\subsection*{Pattern Simulation}

Electron diffraction patterns were simulated for each target zone axis for each crystal using Bloch wave calculations implemented in py4DSTEM \citep{savitzky2021py4dstem, ophus2022automated, Zeltmann2023blochwave, deGraef2003introduction}. Each pattern was simulated using a beam energy of 300 keV, a maximum scattering angle of 2 \AA{}$^{-1}$, and specimen thicknesses from 2 nm to 200 nm thick with a step size of 2 nm. In the models trained in this work, we used only the 20 nm thickness simulations for a total of 3.6 million unique diffraction patterns.  All thicknesses are contained in the datasets published with this work. 

We simulated our electron diffraction patterns using the Bloch wave method. This method allows the generation of a simulated pattern by calculating the scattering matrix using the Fourier coefficients of the crystal electrostatic potential, the beam direction, and the sample thickness \citep{Zeltmann2023blochwave}. This allows the accurate generation of a simulated pattern for an arbitrary crystal tilt and specimen thickness. The output of each Bloch wave simulation is an $(n,3)$-sized set of $n$ Bragg peaks, each defined by a vector ($q_x$,$q_y$,$I$), where $q_x$ and $q_y$ are the x and y position of the peak in diffraction space and $I$ is its intensity. This format is equivalent to a diffraction peak representation after Bragg disk detection as defined by Savitzky et al.~
\citep{savitzky2021py4dstem}. As the number of spots present in each pattern can vary widely, we performed a preprocessing step to create a uniformly sized input to our ML models, which is a requirement for the architecture we've chosen. We decomposed each pattern into a vector of complex-valued radial representations. Each entry in this vector is defined by $\tau_{m,p}$ where $m$ and $p$ are the radial and angular orders, respectively. Each radial function $m$ has a width of 0.05\AA{}$^{-1}$ and a number of angular modes defined by $p$. $\tau_{m,p}$ is given by

\begin{equation}
    \tau_{m,p} = 
    \sum_{j} \beta_{m,p}(\vec{q}) I_{j}(\vec{q}) 
\end{equation}

where the sum is evaluated over all diffraction spots contained within the radial function, I$_j$ is the intensity of the corresponding spot, and $\beta_{m,p}$ is defined in equation 2 below:  

\begin{equation}
    \beta_{m,p}(\vec{q}) = 
    \max \left(
        1 - \frac{||q| - m \Delta q|}{\Delta q},
        0
    \right)
    \exp \left(
        i \, p \, \phi_q
    \right),
\end{equation}
where $|q|=\sqrt{{q_x}^2+{q_y}^2}$ is the magnitude of each scattering vector, $\Delta q$ is a chosen radial step size, $\phi_q=\rm{atan2}(q_y,q_x)$ is scattering vector angle, and $m$ and $p$ are the radial and annular orders respectively. Each bin has a total width of 2$\Delta q$, and overlaps adjacent bins by $\Delta q$. The top row of Figure~\ref{fig_model_outline} shows some examples of our basis functions with 6-fold rotational symmetry, i.e.~$p=6$.

We used 41 radial bins extending to a maximum scattering angle of 2 \AA{}$^{-1}$, with a step size of $\Delta q = $ 0.05 \AA{}$^{-1}$. For each radial bin, we included 13 different angular frequencies corresponding to $p=0, 1,..., 12.$ We selected a maximum angular frequency of 12, corresponding to twice the value of the highest observed symmetry present in our simulated dataset of 6-fold rotational symmetry. We selected the radial step size by balancing higher bin density with memory constraints for model training. This yields a 533-length complex vector for each diffraction pattern, which was converted into the absolute magnitude and complex argument components and stacked yielding a length 1,066 fixed sized real vector as the input to the random forest model. An illustration of this process is shown in Figure \ref{fig_model_outline} and Figure S2.

Although our dataset contains triclinic materials, our modeling procedure struggled significantly to predict them accurately. The crystal system model when triclinic is included is shown in Figure S3. The overall accuracy on individual patterns when triclinic is included is 63\%, a decline from 67\% when triclinic is not included. Particularly noticeable is the model's insensitivity to positive identification of triclinic systems; only 14\% of triclinic materials in the test set are correctly identified as triclinic. The vast majority of triclinic materials are predicted incorrectly as either monoclinic or orthorhombic. Since triclinic materials make up only slightly less of the dataset than trigonal, which can be predicted successfully, it can likely be ruled out that the model is uniformly biased against triclinic due to a low volume of triclinic training data (Figure S1). Rather, in this case it is possible that many of these triclinic materials are only very slightly distorted from monoclinic structures, and therefore the model has difficulty distinguishing between them. Additionally, it is also possible that the information contained in the 2D diffraction patterns is insufficient to differentiate the 3D differences between monoclinic and triclinic structures. Due to this observation, and the lack of triclinic materials in most materials science works, we have removed them from the model presented in this work. This resulted in a final training and validation dataset of $\approx$3.4 million patterns.

\subsection*{Random Forest Modeling}

In this work we used random forest (RF) models for crystal structure prediction. These RF models were trained using Scikit-learn's RandomForestRegressor and RandomForestClassifer models \citep{sklearn_source}. We trained 13 distinct random forest models to develop our full architecture. These were: a model to predict the crystal system, 6 models to predict space group, and 6 models to predict the lattice parameter lengths, \(a, b, c\). Each of these 6 models correspond to a crystal system, excepting triclinic, and they are trained on only materials from that crystal system. See Figure \ref{fig_model_outline} for an illustration of how an arbitrary input passes through our architecture. The hyperparameters of each class of model are shown in table 1. These hyperparameters were selected due to observations that increasing the number of trees and the max depth beyond the values shown in table 1 led to minimal improvement in model accuracy, and led to large increases in the memory size of the models. This set of hyperparameters optimizes model performance and memory constraints, where our goal was to keep each model less than 10 GB.

\begin{center}
\begin{tabular}{ c c c c}
 Parameter & Crystal & Lattice Constant & Space Group \\ 
 \hline
 Num Trees & 80 & 80 & 128 \\ 
 Max Depth & 40 & 100 & 30 \\  
 Max Features & all & all & all    
\end{tabular}
\end{center}

The dataset was split into train and test components using a 75/25 random train/test split function from Scikit-learn, resulting in 2.5 million patterns for training and 0.9 million for testing. One caveat to this approach was that the random split was done along materials, such that all simulated patterns from a given material were assigned to either the train or test group. This was done in order to remove potential model biases due to the model receiving patterns from the same material, but different zone axes, in the training and test set. This ensures that the test set was comprised entirely of materials absent from the training set. The set of training and testing materials was preserved across the submodels as well, ensuring that test set materials in the crystal system model were also test set materials for each of the lattice constant/space group models. This allows the validation of the entire architecture using our test set, rather than considering individual components separately. 

The structure of this model allows for the input of either individual or multiple patterns. In the case of an individual pattern, the prediction is returned as well as the predictions for each of the decision trees comprising the random forest, which can be thought of as the model's internal confidence in its prediction. In the case of a categorical prediction (crystal system, space group) the most commonly predicted label across the decision trees is assigned as the prediction. In the case of a regression task (lattice parameters) the median value of the decision trees is assigned as the prediction.  

\begin{figure*}[htbp]
    \centering
    \includegraphics[width=18cm]{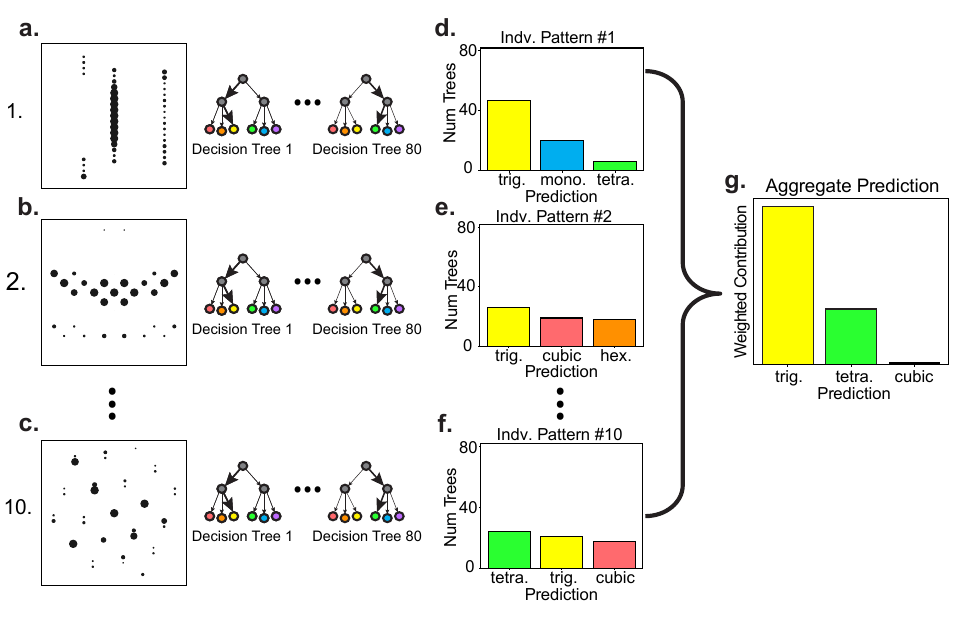}
    \caption{How aggregate predictions are generated from a set of individual electron diffraction patterns from the same material. Figure 3a-c shows a subset of the raw individual patterns, d-f shows the predictions on those patterns and g shows the resulting aggregate prediction. The true crystal system for this material is trigonal.}
    \label{fig_aggregation_illustration} %This is for you to refer to the figure in maintext text
\end{figure*}

When the architecture is provided multiple patterns, the predictions of each individual pattern are aggregated to generate an ensemble prediction. In the case of a categorical prediction (crystal system, space group) the aggregation is done by a weighted average of the individual predictions. The weight attached to each individual prediction is the number of decision trees predicting the most common label, subtracted from the second most common label and divided by the number of decision trees, in this case 80 for crystal system prediction and 128 for space group prediction. Using this procedure, termed difference aggregation in this work, the confidence of each individual prediction is leveraged to determine the aggregate prediction. The confidence in the aggregate prediction is the weighted sum of the predicted label divided by the sum of all the weights across all the individual pattern predictions. See Figure \ref{fig_aggregation_illustration} for an illustration of how an aggregate prediction is generated from individual predictions for crystal system inference. For the aggregate lattice parameter prediction, the median value of the predictions of each individual pattern is returned as the prediction. 

The architecture of our crystal structure prediction model flows as follows. First, the pattern(s) is(are) decomposed into our radial basis representation. Second, the radial representation of each pattern is passed into the RF model which predicts the crystal system. This model returns a prediction and the confidence in that prediction. Based on the predicted crystal system, a submodel is then selected for the prediction of the space group and lattice parameters. Each of the 6 crystal systems used in this work has a unique model for this purpose. The pattern(s) is(are) then input into these models, which return values for the lattice parameters and space groups. 

\subsection*{Experimental Electron Diffraction Patterns}
The performance of our random forest models on experimental data was validated using a 4D-STEM scan of a sample of gold nanoparticles (AuNPs) on a carbon support film. The 4D-STEM data was acquired on the on the TEAM I microscope, operated at 300kV with a probe convergence angle of 0.63 mrad and a step size of 2.1 nm. The diffraction patterns were recorded on the Dectris Arina camera with a dose of approximately 400 e$^-$/\AA$^2$. To compare the input experimental data to our simulated data, we utilize principal component analysis (PCA) to decompose the input vectors for our experimental data and our simulated data corresponding to FCC Au into a two dimensional representation. This is shown in Figure S4, which demonstrates that the experimental data contains many patterns that are similar in scope to our simulated FCC Au electron diffraction patterns, and also many that are significantly different from the simulated Au, demonstrating the versatility of this model on data not directly encompassed by the simulated data.

Upon pattern acquisition, electron diffraction patterns are processed to lists of Bragg peaks from the raw experimental pattern using py4DSTEM \citep{savitzky2021py4dstem}. The resulting datacube contains acquired Bragg lists at each probe position on a 256x256 grid scan. This array of patterns is filtered to remove Bragg lists containing fewer than 5 diffraction spots. The filtered dataset is then transformed into our radial representation and the crystal system is predicted with our random forest model. The resulting array of crystal system predictions is median filtered in a 5x5 grid and pixels with a difference confidence lower than 0.005 post median filtering are removed, as this reflects a very low gap between the number of trees predicting the plurality crystal system vs the second most commonly predicted. Two possible avenues are followed after crystal system determination: lattice prediction of all probe positions with a median filtered crystal system prediction using either one lattice prediction model or the lattice prediction model corresponding to the crystal system prediction at that probe position. Both avenues are utilized in this work: lattice parameter prediction using only the cubic model, due to the fact that with a-priori knowledge we know that FCC Au is cubic, and lattice parameter prediction using the model corresponding to the crystal system prediction of the individual pattern at each probe position, simulating a situation where we have no prior information about the crystal system. 

\section*{Results and Discussion}

\subsection*{Crystal System Accuracy}
\begin{figure*}[ht]
    \centering
    \includegraphics[width=18cm]{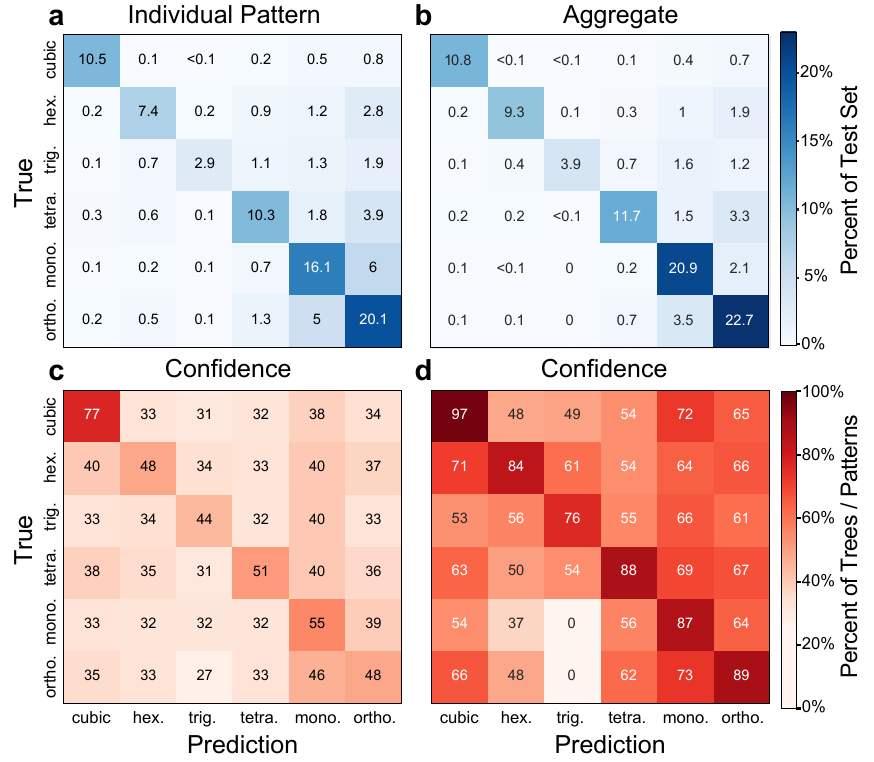}
    \caption{Prediction accuracy and confidences for individual patterns and aggregate predictions. The values in (a) and (b) are the percentage of the test set representing that prediction/true pairing. (a) shows the predictions on individual patterns while (c) shows their prediction confidences, calculated by determining the percentage of decision trees in the random forest model corresponding to the predicted crystal system. (b) shows the predictions on aggregate patterns and (d) shows the corresponding confidences, calculated by determining the percentage the weighted aggregation values corresponding to the predicted crystal system.}
    \label{fig_confusion_matricies} %This is for you to refer to the figure in maintext text
\end{figure*}

The overall accuracy of our crystal system prediction model is 67\% when tested on 859,656 patterns from 8,549 previously unseen materials. A confusion matrix of the performance of the model on the six crystal systems is shown in Figure \ref{fig_confusion_matricies}a. The on-diagonals correspond to correct predictions, and the row and column errors are the false negative and false positive predictions, respectively. The values are the percentage of the test dataset attributed to the prediction. The largest source of error is orthorhombic and monoclinic crystals incorrectly classified as each other. This error could stem from the similarity in the crystal systems and the inherent challenge in using 2D diffraction patterns to differentiate the inherently 3D difference between
monoclinic and orthorhombic structures. Orthorhombic crystals also contain the highest false positive prediction rate, 43\% for the individual prediction, as shown in Figure S5. We attribute this to the abundance of orthorhombic crystals in the training dataset, leading to an orthorhombic bias. This seems to be the greatest indicator of inaccuracy, as the false positive rate is strongly correlated with the prevalence of that crystal system in the dataset (Figure S5). Similarly, trigonal materials have few false positives and are the least common material in the dataset, further supporting this supposition. 

Figure \ref{fig_confusion_matricies}c contains average values for the prediction confidence of each of the corresponding entries in Figure \ref{fig_confusion_matricies}a. The confidence value for a single pattern prediction is the percentage of the individual decision trees predicting the crystal system. Figure \ref{fig_confusion_matricies}c shows the prediction confidence reflects the accuracy confusion matrix, where accurate predictions are linked to higher confidence values on average. 

The prediction confidence can function as an internal uncertainty metric and aide in interpreting the model's predictions. Analyzing the accurate predictions shows an average confidence of 55\% when all accurate predictions are averaged together. Additionally, averaging the confidences of the inaccurate predictions shows a decrease in confidence to 38\%, showing accurate predictions are associated with higher confidence on average. 

Although there are misprediction regions of higher confidence, such as monoclinic's misprediction as orthorhombic with the same average confidence with which orthorhombic is correctly predicted, it is possible that minor distortions in the orthorhombic lattice make it nominally monoclinic, but still functionally orthorhombic, and cause the model to render a higher confidence prediction. The uniquely high symmetry of cubic materials is also identified by the model, resulting in accurately identified cubic materials having a prediction confidence over 20 percentage points higher than the next highest crystal system.

\subsection*{Crystal System 
Accuracy of Aggregate Predictions}

The first component of this work produced a crystal system accuracy of 67\% on individual patterns and developed a heuristic for inferring the prediction's accuracy based on the model's internal confidence. However, often many diffraction patterns are taken in an experiment, ideally at multiple orientations. Therefore, we have also developed an architecture to examine predictions across multiple patterns and aggregate them into a single prediction. This approach is particularly useful for situations where the user is varying the pattern acquisition zone axis, either through sample tilting or by sampling diffraction patterns from multiple individual particles that are randomly oriented, and allows a higher accuracy and stronger confidence prediction of the crystal structure. The aggregate prediction is determined by a weighted average of the individual patterns, termed difference aggregation, as described in the methods and shown in Figure \ref{fig_aggregation_illustration}. 

When using the predictions generated from aggregating 10 patterns at different orientations into a unified prediction, the crystal system prediction accuracy increases to 79\%. The explanation for this increase in accuracy can be seen in Figure \ref{fig_aggregation_illustration}. In Figure \ref{fig_aggregation_illustration}, individual predictions of three electron diffraction patterns are shown, and these represent the three main types of predictions rendered by the model trained on individual patterns. The first, Figure \ref{fig_aggregation_illustration}a,d, is an accurate prediction with high confidence, showing that 47 of the 80 decision trees in the random forest model make the same prediction as the most common prediction, in this case trigonal. Using the difference aggregation method, this yields a prediction of trigonal with a weight of (47 - 20)/80 = 0.3375, where 20 is the number of decision trees predicting monoclinic, the second most common prediction. The second type, Figure \ref{fig_aggregation_illustration}b,e, is a correct prediction that occurs with low confidence. In this case, the model returns a prediction of trigonal but only 26 of the 80 decision trees make the same prediction. Additionally, an appreciable 19/80 decision trees make the incorrect prediction of cubic. This leads to a prediction of trigonal with very low weight, (26-19)/80 = 0.0875. The third case is an inaccurate prediction, which also typically occurs with low confidence. This case is illustrated in Figure \ref{fig_aggregation_illustration}c,f, where the model incorrectly assigns the material as tetragonal, but only 24 of the 80 decision trees predict this. Additionally, an almost identical number, 21, predict trigonal, showing an extreme lack of confidence in this prediction. This yields a prediction of tetragonal with a very low weight, (24-21)/80 = 0.0375 . 

The difference aggregation method emphasizes predictions that occur with high confidence, which are often more accurate. Additionally, as can be seen in \ref{fig_aggregation_illustration}c, low confidence predictions occur when the acquisition is quite off zone, making understanding the pattern more difficult for both modeling and human experts.  Figure \ref{fig_aggregation_illustration} shows that by taking this aggregate of 10 patterns on this material, the incorrect predictions are simply lumped into the aggregate, and often contribute very little weight to the aggregation. This produces a high confidence prediction of trigonal, where its weighted contribution is roughly 3 times that of the next highest, tetragonal. The confusion matrix shown in Figure \ref{fig_confusion_matricies}b shows the results of running our entire test set through this architecture, where 10 patterns across different crystal orientations are each predicted individually and then aggregated to serve as a final prediction for that material. The crystal system accuracy improves by 12 percentage points in the aggregate case, increasing from 67\% to 79\%. The highest percent increases come from trigonal and hexagonal, which improve by 34\% and 26\%, respectively. The lowest comes from cubic, which improves by only 3\%. This, however, is likely a reflection of how highly accurate the cubic prediction was on the individual case. The confusion between orthorhombic and monoclinic, and the general overprediction of orthorhombic, has a lower prevalence in the aggregate case. The greatest improvement comes from the inaccurate prediction of monoclinic when the true value is orthorhombic, which falls from 6\% of the test set to only 2.1\% in the aggregate case.

Figure \ref{fig_confusion_matricies}d shows the confidences of the aggregate predictions, which is defined as the percentage of the weighted sums making that prediction, not to be confused with the percentage of individual decision trees which determined model confidence for the predictions on individual patterns in Figure \ref{fig_confusion_matricies}c. As with the individual confidences, the highest values for the prediction confidence in Figure \ref{fig_confusion_matricies}d run along the diagonal of the confidence confusion matrix. However, in comparison to Figure \ref{fig_confusion_matricies}b the diagonal stands out as higher confidence than the inaccurate predictions, particularly around orthorhombic and monoclinic confusions, which was an issue in the individual case. A simple average of the confidence of every material correctly predicted gives an average confidence of 88\%. A similar procedure for the inaccurate predictions shows the average confidence falls to 66\%. 

Additionally, examining a column of the confidence confusion matrix, as in the individual case in Figure \ref{fig_confusion_matricies}c, shows the highest value is the diagonal value, and often by a margin greater than 10. Therefore, setting a crystal system specific threshold prediction confidence for a prediction to be considered trustworthy may allow the systematic determination of accuracy based on prediction confidence. This idea is further examined in Figure S6, which visualizes the frequency of accurate vs inaccurate predictions having high and low confidence. For the cubic, hexagonal, tetragonal and trigonal predictions, a higher confidence prediction is visibly linked to a greater change of accuracy in the crystal system prediction for individual pattern predictions and aggregate predictions (Figure S6). The high number of materials and patterns in the test set makes it possible to determine the likelihood a prediction is correct based on its confidence. For example, an individual pattern hexagonal prediction with a confidence of between 0.3 and 0.35 has only roughly a 66\% chance of being correct, while a hexagonal prediction with a confidence of between 0.5 and 0.55 is correct over 90\% of the time. The confidence is a less useful metric in the individual pattern case for orthorhombic and monoclinic, given that the correct predictions are often rendered with low confidence. In the 10 pattern aggregate for these crystal systems this is mitigated somewhat by a higher density of high confidence predictions, and the confidence is much more strongly linked with accuracy than in the individual pattern case.

\subsection*{Lattice Parameter and Space Group Accuracy}

Upon determination of the crystal system, each pattern is then input into the corresponding lattice constant/space group model. We trained 12 models to accomplish this task, and these 12 models are comprised of 6 sets of 2, with each set of 2 corresponding to lattice parameter and space group prediction models for each of the 6 crystal systems studied. For example, as seen in Figure \ref{fig_model_outline}, once trigonal is predicted as the crystal system, the model then flows into the space group and lattice parameter prediction models trained specifically on trigonal materials. 

\begin{figure*}[ht]
    \centering
    \includegraphics[width=18cm]{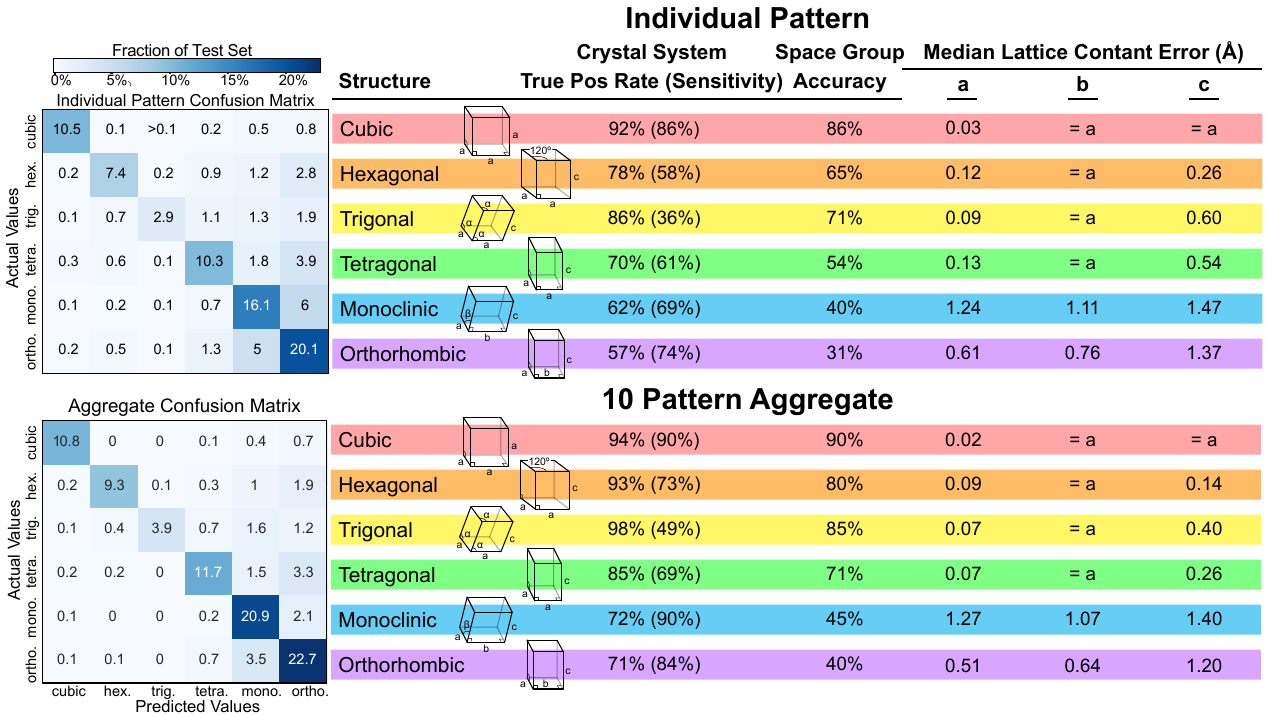}
    \caption{Summary results for the full model architecture. In the crystal system column, the True Pos Rate is the percentage of correct predictions made of that crystal system. For example, in the individual case 92\% of cubic predictions are correct. The other 8\% are incorrectly called cubic. The Sensitivity denotes the percentage of the total diffraction patterns within that crystal system correctly identified by the model. For example, in the individual case 86\% of cubic materials are labeled cubic, while 14\% are labeled as some other crystal system. The space group accuracy is the simple percentage correctly predicted, and the lattice constant error is the median error of the patterns predicted to correspond to that crystal system when they're inputted into the lattice prediction model. Both lattice parameter and space group prediction accuracies include materials incorrectly assigned to that crystal system.}
    \label{fig_model_accuracy} %This is for you to refer to the figure in maintext text
\end{figure*}

\par Figure \ref{fig_model_accuracy} shows the accuracy of the lattice and space group models on each crystal system when using the lattice/space group model matching the pattern's predicted crystal system, regardless of whether this prediction is correct. It is important to note how this impacts the topline accuracies, as we are including situations where a pattern's crystal system has been mispredicted. In the case of space groups, the result is straightforward as, for example, a space group prediction using the hexagonal model on a trigonal material mislabeled hexagonal is guaranteed to be incorrect. The hexagonal model will still return a space group prediction, however, it will only predict from hexagonal space groups. 

\par
\begin{figure*}[ht]
    \centering
    \includegraphics[width=18cm]{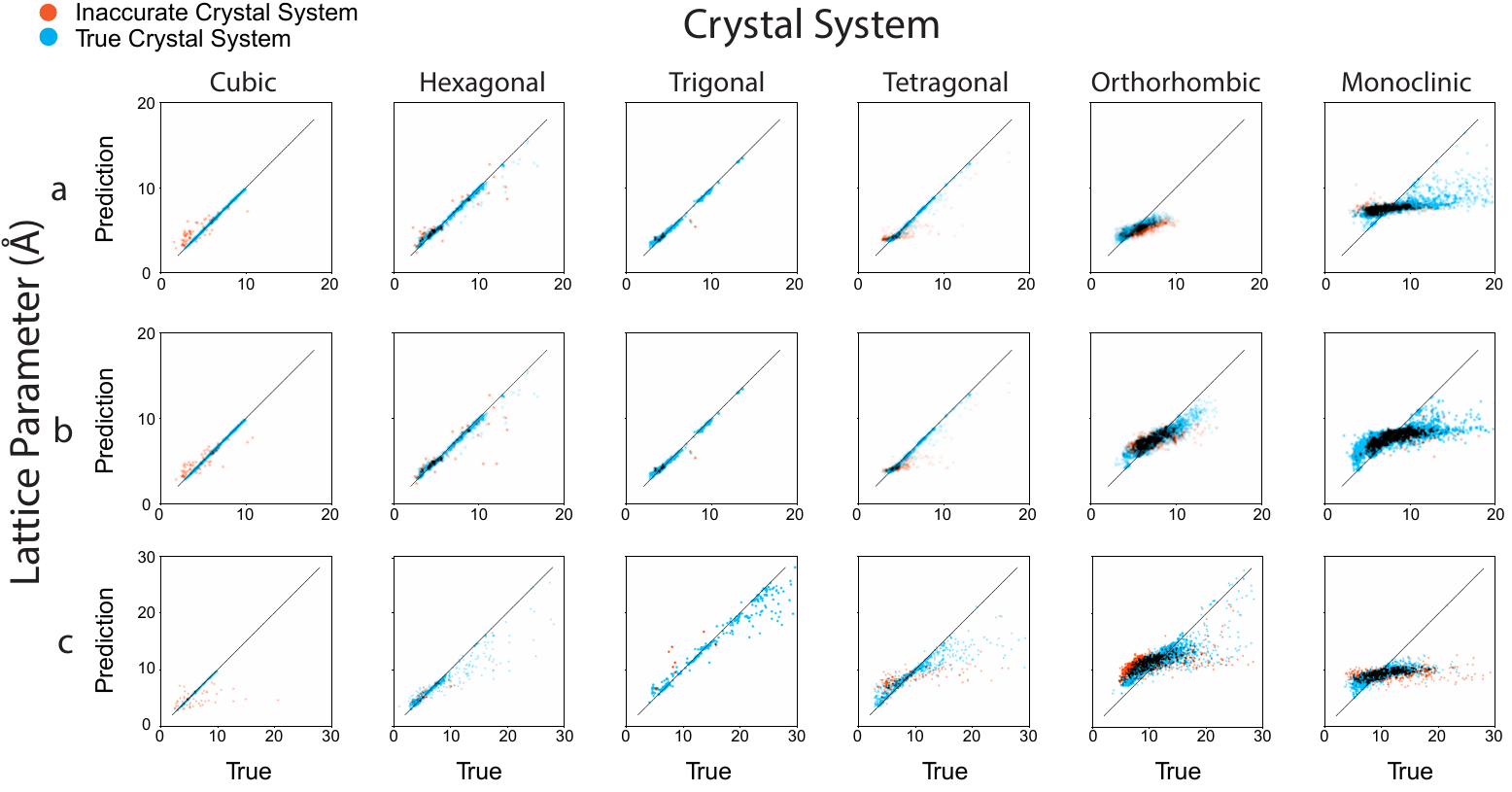}
     \caption{2D histograms showing the accuracy of the 10 pattern aggregate lattice parameter prediction on materials predicted to correspond to each crystal system. The orange colors denote mispredictions in the crystal system while the blue shows correct predictions. Increasing color density shows more measurements at that location. The diagonal lines on each plot indicate the location of perfect predictions.}
     \label{fig_lattice_accuracy_aggregate} 
\end{figure*}

 The lattice parameter prediction, however, presents a more complex case since the lattice prediction can still be reasonably accurate even if the crystal system has been misidentified. This is especially true for boundary cases between crystal systems. For example, if a crystal with very similar, but not exactly the same, values for $a$, $b$, and $c$ is called a cubic by the crystal system prediction model, the cubic model may return a reasonable lattice prediction. Figure \ref{fig_lattice_accuracy_aggregate} and Figures S7-S9 show the inaccurate crystal system identifications result in lattice parameter predictions that are less accurate on average, although they occasionally return highly accurate predictions.     

\par The median prediction error was chosen to represent the accuracy of the lattice prediction models due to the high degree of skew found in histograms of the absolute error in virtually every crystal system and lattice constant (Figures S8 and S9). The individual pattern prediction is taken from the median prediction of the 80 trees in the random forest model, as this was shown to perform far better than the mean value. The predicted values for the 10 pattern aggregate prediction are determined by taking the median result across the 10 patterns. In the cubic case, shown in the red row in Figure \ref{fig_model_accuracy}, $a=b=c$ and therefore the median error is the same for each parameter, with a median absolute error of 0.03 \AA{} and 0.02 \AA{} for the individual and aggregate cases, respectively. Since cubic materials are also identified with high accuracy from the crystal system prediction, this model can produce a very detailed and accurate quantitative prediction of the crystal structure for a cubic material.

\par Tetragonal, hexagonal and trigonal materials follow a similar pattern in Figures \ref{fig_model_accuracy} and \ref{fig_lattice_accuracy_aggregate}, where, since $a=b$, both of these parameters are predicted with a median absolute error of around 0.1 \AA{} for the individual case, with a consistent improvement of roughly 33\% when switching to the aggregate case. The c axis, however, is harder to identify, and the median absolute error for this parameter is higher and more variable across these three crystal systems. Hexagonal materials retain a reasonable accuracy with a median absolute error of 0.26 \AA{} and 0.14 \AA{} for individual and aggregate patterns respectively. Trigonal and tetragonal are noticeably worse for the individual case, both between 0.5 \AA{} and 0.6 \AA{}. However, tetragonal improves far more in the aggregate case, dropping to 0.26 \AA{} while trigonal remains at 0.40 \AA{}. This higher error is likely due to the higher variability in $c$ across these materials, as well as the presence of a few very high $c$ values in the simulated dataset which introduces more uncertainty into the prediction and causes a few large outliers. The higher spread of high $c$ values is particularly noticeable in the trigonal materials in Figure \ref{fig_lattice_accuracy_aggregate} and Figure S7, but is clearly defined in all three of these systems.

Additionally, higher values of $c$ present a challenge for our radial basis representation. One possible explanation comes from the underlying physics of a material with a long c axis and how this is reflected in electron diffraction. Such materials produce patterns with spots very close to the origin and very close to each other, as a long c axis in real space corresponds to very close spacing in reciprocal space. In the case of an unusually long c axis, such as 30 \AA{}, this produces spots 0.03 \AA{}$^{-1}$ apart, which can be challenging for our radial basis representation, which uses radial bins currently set to have a $\Delta q$ of 0.05 \AA{}$^{-1}$ (see equations 1 and 2). Model performance on the c axis begins to deteriorate slightly after 10 \AA{}, and becomes far more pronounced after 20 \AA{}, as shown in Figure \ref{fig_lattice_accuracy_aggregate} and Figure S7. This is to be expected given sampling frequency of our basis set, as the Nyquist frequency, or the location where a decomposition begins to struggle to regenerate the underlying distribution, begins at half the resolution of the decomposition basis. In our radial basis, this is 0.1 \AA{}$^{-1}$, corresponding to a lattice parameter of 10 \AA{}. Another likely explanation is the low representation of long c axis materials in the training set. This makes the model unwilling to predict high values for any lattice constant. Additionally, as many experiments are unlikely to feature materials with unit cell parameters of greater than 20 \AA{}, this is unlikely to be a major detriment to this model's applicability to most experimental electron diffraction patterns. 

The introduction of different values for $a$, $b$, and $c$ in orthorhombic materials results in a decline in model accuracy across all three parameters, likely stemming from the difficulty in extracting predictions of a unit cell with three unique dimensions from two-dimensional input. The degeneracy in the 2D electron diffraction pattern created by three different unit cell lengths makes deducing the lattice parameter values for orthorhombic very challenging, and even the moderate degree of correlation between the predicted and true lattice constants found in the orthorhombic plots in Figure 6 is surprising from a fundamental perspective. Across $a$, $b$, and $c$ axes a similar pattern exists as the $c$ predictions of tetragonal, trigonal and hexagonal, where the lower valued lattice parameters are predicted with reasonable accuracy while some of the larger ones open up more significant errors, which are commonly underestimates (Figure \ref{fig_lattice_accuracy_aggregate} and Figure S7. Adding further dimensions to the lattice in the monoclinic crystal system renders the model almost completely unable to yield any meaningful lattice information, which is likely a physical limitation of working with two-dimensional data. The struggles with monoclinic which, due to the fourth degree of freedom, is noticeably worse than orthorhombic, likely also stems from correlations between the errors, i.e. if the model is unable to predict the a axis correctly it will likely fail on b and c as well. This is noticeably different from the orthorhombic case, which is able to deliver reasonable predictions on the shorter axes. This accurate prediction of some component of the unit cell may allow the model to use these predictions to make an inference about the longer axes. A possible future direction to mitigate these issues is to train this model using PED data, which will encode 3D information into the diffraction pattern and will be better able to provide information on a 3D system \citep{Gemmi2019Electron}.  

The space group prediction models are able to deliver nearly perfect accuracy on cubic materials, correctly identifying the space group almost 90\% of the time on individual patterns, with an improvement of a few percentage points in the aggregate case. Hexagonal, trigonal and tetragonal retain a reasonable accuracy of  50-70\% on individual patterns while improving to 70-90\% in the aggregate case. Monoclinic and orthorhombic begin to breakdown somewhat, mainly due to the high number of false positives in monoclinic and orthorhombic. These false positives introduce a large set of patterns that the model is guaranteed to mispredict. This is a relatively minor issue in cubic, hexagonal, trigonal and tetragonal, with tetragonal being the highest false positive rate at 15\% in the aggregate case. However, in the 10 pattern aggregate case for monoclinic and orthorhombic, 28\% and 29\% of predictions are false positives, respectively. The overall space group accuracy is 45\% and 40\% for each of these crystal systems, including predictions on misclassified crystal systems. Therefore, using a conditional probability, we can deduce the monoclinic/orthorhombic accuracy conditional on correctly predicting the crystal system. When the initial crystal system is predicted correctly, the space group accuracy is 63\% and 56\% for monoclinic and orthorhombic, respectively. Improving the crystal system prediction for these crystal systems or having a-priori knowledge of the crystal system for an experimental test sample will lead to a significant increase in space group accuracy. As noted above in the lattice prediction discussion, a possible future direction to improve crystal system accuracy is to train this model using PED data. Precession better preserves the intensity-ordering and forbidden reflections inherent to the space group, which will be better able to provide information on a 3D system, and should significantly enhance crystal system identification and space group accuracy. This is an active area of future work, and precession electron diffraction simulations are currently being added to py4DSTEM for this purpose.

\subsection*{Experimental Validation}
\begin{figure*}[ht]
    \centering
    \includegraphics[width=12cm]{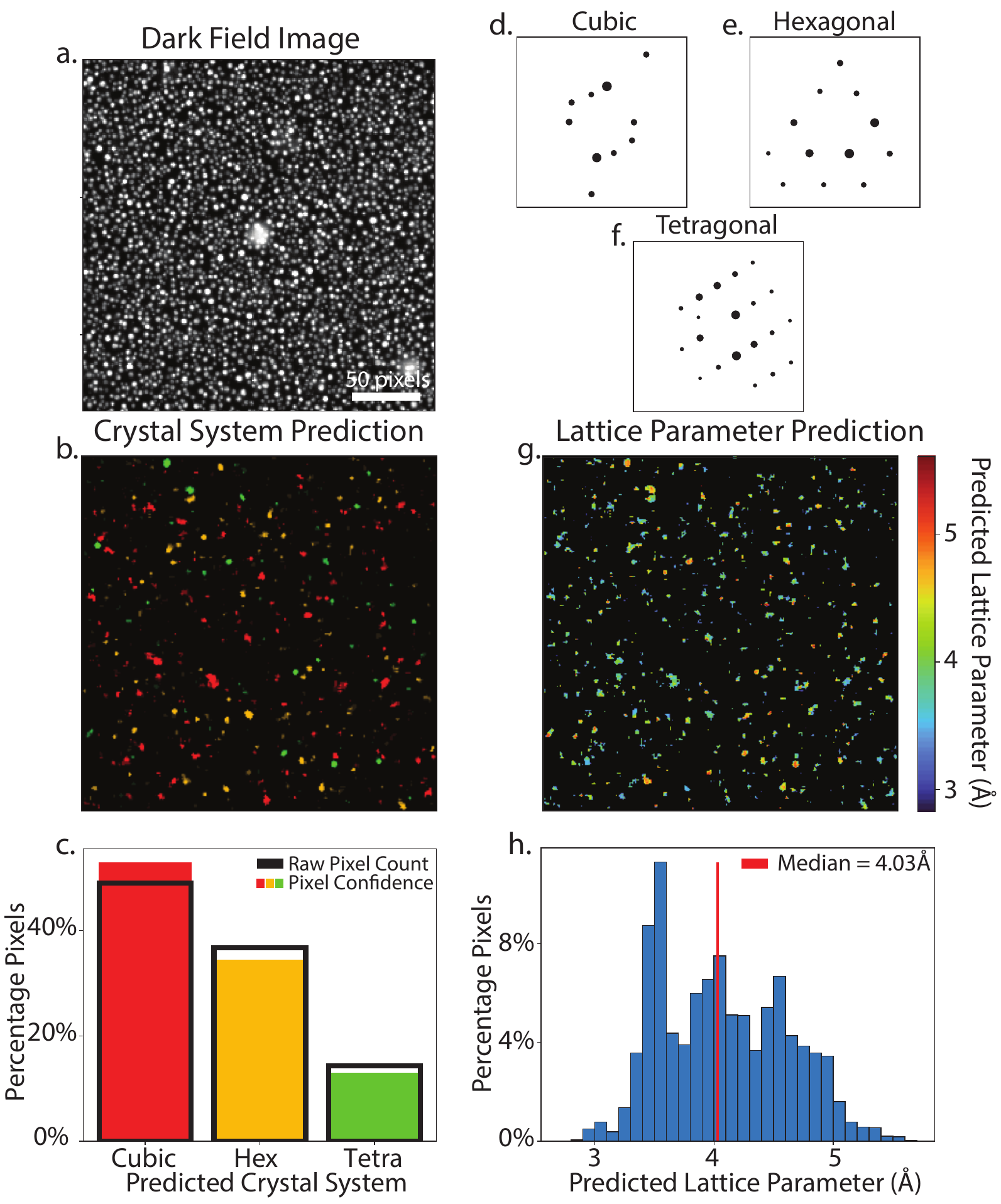}
     \caption{Application of the crystal structure prediction model on a set of experimental electron diffraction patterns taken from a sample of AuNPs. (a) the dark field image of the scan. (b-c) the crystal system prediction where red corresponds to cubic, orange to hexagonal and green to tetragonal predictions. The color brightness in (b) is correlated to the prediction confidence, with brighter pixels having a higher prediction confidence. The black rectangle in (c) corresponds to the raw pixel counts corresponding to each crystal system and the bars indicate the counts weighted by prediction difference confidence. (d-f) sample electron diffraction patterns predicted each of the three returned crystal systems. (g-h) the lattice parameter prediction where (g) are the same pixels predicted in (b).}
     \label{fig_experimental_application} 
\end{figure*}

To test the accuracy of this model on experimental data we applied our architecture to a 4D-STEM scan of gold nanoparticles (AuNPs). The 4D-STEM scan was filtered to ensure predictions weren't returned on probe positions where crystals weren't present and to remove low confidence predictions. This was done by removing probe positions where there were fewer than 5 diffraction spots, median filtering the results in a 5x5 grid and removing pixels with a difference confidence lower than 0.005 post median filtering, as this reflects a very low gap between the number of trees predicting the plurality crystal system vs the second most commonly predicted. The results are shown in Figure \ref{fig_experimental_application}.  

Figure \ref{fig_experimental_application}b and c show the model can accurately identify around half the AuNP pixels as cubic, and when weighting the predictions by prediction confidence we see the cubic predictions are slightly more confident than the hexagonal and tetragonal predictions, on average (Figure \ref{fig_experimental_application}c). 

While many crystals in Figure \ref{fig_experimental_application} are correctly identified as cubic, it is clear that an appreciable amount are also predicted to be hexagonal, and an additional smaller amount are predicted tetragonal. The mispredictions of hexagonal and tetragonal can be explained by the existence of multiple orientations for FCC Au, such as the 111 axis, where the electron diffraction pattern is identical to that of a hexagonal pattern \citep{Ponce2021advances, Zachman2022measuring}. Therefore, it would be beyond the scope of this model to return a prediction of cubic for such an electron diffraction pattern. This is further demonstrated in Figure \ref{fig_experimental_application}d-f, which shows extracted electron diffraction patterns that are predicted cubic, hexagonal and tetragonal, respectively. Furthermore, the propensity for small AuNPs to form multiply-twinned structures means that many diffraction patterns may be integrating more than one twin species, which is a feature not present in the training data for our model \citep{Marks1983_HREM}.  

When the prediction type is shifted from a per pixel prediction to a per particle prediction, done by isolating and predicting the brightest diffraction pattern in each particle, the prediction spread falls to a roughly 40:40:20 cubic:hexagonal:tetragonal ratio. This is likely due to the issues outlined above, and additionally shows a preference for the [111] zone axis, as this axis appears to regularly show up as the brightest pattern in these crystals, while aggregating over the particle as a whole yields more particles labeled cubic. This can be explained by the preferential orientation of AuNPs being in the 111 direction 
\citep{Mateo2016111oriented}. Additionally, previous work predicting crystal structure of AuNP samples from electron diffraction patterns found an almost identical spread of predictions between hexagonal and cubic as presented in Figure \ref{fig_experimental_application}c, showing that degeneracy in FCC cubic AuNPs is a common problem across AuNP samples 
\citep{Ponce2021advances}. 

An additional utility of this model is the prediction of the lattice constant. These results are shown in Figure \ref{fig_experimental_application}g-h. Figure \ref{fig_experimental_application}g shows the pixels predicted in Figure \ref{fig_experimental_application}b colored by their lattice constant prediction rather than their crystal system prediction. Figure \ref{fig_experimental_application}h shows that, despite the rather large spread of per pixel predictions, the median lattice constant prediction is 4.03 \AA{}, very close to the true lattice constant for Au of 4.08 \AA{} 
\citep{Guan2009relaxation}. Given the high standard deviation of predictions, the exactness of this result shouldn't be interpreted as experimental accuracy to 0.05 \AA{}, but it does show a high degree of accuracy in the aggregate case of lattice parameter prediction. For the results shown in Figure \ref{fig_experimental_application}g-h, all patterns were predicted with the cubic lattice parameter model. The spread using the lattice parameter model corresponding to each individual pixel's crystal system prediction is shown in Figure S10. Figure S10 shows that using the hexagonal/tetragonal model for patterns predicted hexagonal/tetragonal leads to a series of predictions where the $c$ axis is predicted slightly higher than $a$ and $b$, resulting in a mild underestimation of the $a$ and $b$ axis and a slightly larger overestimation of the $c$ axis. These results are consistent with situations where patterns from the incorrect crystal system are predicted using a different lattice parameter model (Figure \ref{fig_lattice_accuracy_aggregate} and Figure S7).  

One existing experimental limitation is the inability to predict patterns comprised of multiple grains \citep{Marks1983_HREM}. This is because the training data for the model was comprised only of diffraction patterns from individual grains. Therefore, the addition of patterns comprised of multiple grains appears as additional unexpected diffraction spots which the model is then unable to separate into component patterns. This results in the model typically assigning these patterns almost exclusively as orthorhombic or monoclinic, as it attempts to determine what individual crystal could produce the multiple grain pattern. In instances where this model is returning a higher than expected volume of orthorhombic or monoclinic predictions, the inputted patterns may be comprised of multiple grains. Future work on this model will augment the training data to include patterns with multiple grains, to ensure increased robustness of this workflow.

\section*{Conclusion}

% the crystal system identification model is attached to methods to determine the uncertainty in the model's prediction, whether one or several patterns are used as input. 

In this work, we simulated dynamical electron diffraction patterns of 20 nm thickness, and these simulated patterns were used to train a hierarchical architecture of random forest models to identify crystal system, space group and lattice constants from unlabeled electron diffraction patterns. This hierarchical model first predicts the crystal system and then transfers the pattern into a crystal system specific submodel to predict space group and lattice parameters. The crystal system model returns an accuracy of 67\% when predicting individual patterns and 79\% when predictions of 10 patterns on different zone axes are aggregated together using a difference confidence aggregation metric. Additionally, we predict the uncertainty in model predictions for single and multiple pattern inputs. This uncertainty is correlated with the accuracy of the prediction. The space group prediction has an accuracy of $\approx$70-90 \% for cubic, hexagonal, trigonal and tetragonal crystal systems, while being less reliable on orthorhombic and monoclinic systems. The lattice prediction varies from median errors of $\approx$0.01-0.5 \AA{} for cubic, hexagonal, trigonal and tetragonal crystal systems, while ranging from $\approx$0.5-1.5 \AA{} for monoclinic and orthorhombic. The main difficulty in the prediction of crystal structure in general, and particularly the lattice prediction of orthorhombic and monoclinic, is the inability to extract complete 3D information from 2D data. Future work will mitigate this issue by training the model on precession electron diffraction data, which encodes 3D information into the patterns and better provides information on a 3D system. This also explains the increase in accuracy when aggregating patterns taken at multiple zone axes. We applied this architecture to a 4D-STEM scan of gold nanoparticles, demonstrating our ability to accurately predict crystal structure and lattice constants. This architecture can be used to significantly accelerate diffraction pattern analysis, especially for situations where the crystal structure is unknown or has not been previously observed. Additionally, integration of this model with a 4D-STEM experiment would allow on the fly labeling of an entire sample with crystal information and detection of outliers. Beyond the construction of this model architecture, the included diffraction dataset of $\approx$360 million diffraction patterns will benefit the scientific community by providing a robust training set for future data driven and automated analysis studies.

\section*{Author Contributions} 
SPG generated the simulated dataset, conducted the machine learning training and analysis and wrote the manuscript. AR provided training and expertise necessary to generate the simulated dataset and developed the technical implementation of the radial basis pattern representation. SMR acquired the experimental data. SZ developed the block wave calculations necessary to simulate dynamical electron diffraction patterns. BS helped develop the simulated pattern workflow. MH provided the expertise and training necessary to use high performance computing for the simulated pattern generation. JC provided experimental electron diffraction expertise and project advising. CO provided experimental electron diffraction expertise, led the collaboration and designed the scope of this work. All authors read, edited and approved the final manuscript.

\section*{Data and Code Availability}
The code base for this work, and links to the dataset, models, and model outputs necessary to reproduce the figures and results shown here can be found at https://github.com/smglsn12/EDiffCrystals. This repository is currently private, but will be shared upon request and will be made public upon publication of this manuscript. 

\section*{Acknowledgments}
This work was primarily funded by the US Department of Energy in the program “4D Camera Distillery: From Massive Electron Microscopy Scattering Data to Useful Information with AI/ML.” Work at the Molecular Foundry was supported by the Office of Science, Office of Basic Energy Sciences, of the U.S. Department of Energy under Contract No. DE-AC02-05CH11231. CO and SMR acknowledge support from the US Department of Energy Early Career Research Program.

\section*{References}
%aapmrev4-2.bst 2019-01-14 (MD) hand-edited version of aapmrev4-1.bst

% \bibliographystyle{MandM}
% \bibliography{refs}    

\end{document}

\title{Supporting Information For Random Forest Prediction of Crystal Structure from Electron Diffraction Patterns Incorporating Multiple Scattering}

\author{Samuel P. Gleason}
%\email{smglsn12@berkeley.edu}
\affiliation{National Center for Electron Microscopy, Molecular Foundry, Lawrence Berkeley National Laboratory, 1 Cyclotron Road, Berkeley, CA, USA, 94720}
\affiliation{Department of Chemistry, University of California, Berkeley, CA, USA}

\author{Alexander Rakowski}
% \email{ARakowski@lbl.gov }
\affiliation{National Center for Electron Microscopy, Molecular Foundry, Lawrence Berkeley National Laboratory, 1 Cyclotron Road, Berkeley, CA, USA, 94720}

\author{Stephanie M. Ribet}
\affiliation{National Center for Electron Microscopy, Molecular Foundry, Lawrence Berkeley National Laboratory, 1 Cyclotron Road, Berkeley, CA, USA, 94720}

\author{Steven E. Zeltmann}
% \email{}
\affiliation{School of Applied and Engineering Physics, Cornell University, Ithaca, NY, 14853, USA}
\affiliation{PARADIM, Materials Science \& Engineering Department, Cornell University, Ithaca, NY, 14853, USA}

\author{Benjamin H. Savitzky}
% \email{}
\affiliation{National Center for Electron Microscopy, Molecular Foundry, Lawrence Berkeley National Laboratory, 1 Cyclotron Road, Berkeley, CA, USA, 94720}

\author{Matthew Henderson}
% \email{}
\affiliation{National Center for Electron Microscopy, Molecular Foundry, Lawrence Berkeley National Laboratory, 1 Cyclotron Road, Berkeley, CA, USA, 94720}

\author{Jim Ciston}
% \email{jciston@lbl.gov}
\affiliation{National Center for Electron Microscopy, Molecular Foundry, Lawrence Berkeley National Laboratory, 1 Cyclotron Road, Berkeley, CA, USA, 94720}

\author{Colin Ophus}
%\email{clophus@lbl.gov}
\affiliation{National Center for Electron Microscopy, Molecular Foundry, Lawrence Berkeley National Laboratory, 1 Cyclotron Road, Berkeley, CA, USA, 94720}

\date{\today}

\maketitle

\bibliography{refs}    

\setcounter{figure}{0}
\renewcommand{\thefigure}{S\arabic{figure}}

\begin{figure*}[ht]
    \centering
    \includegraphics[width=9cm]{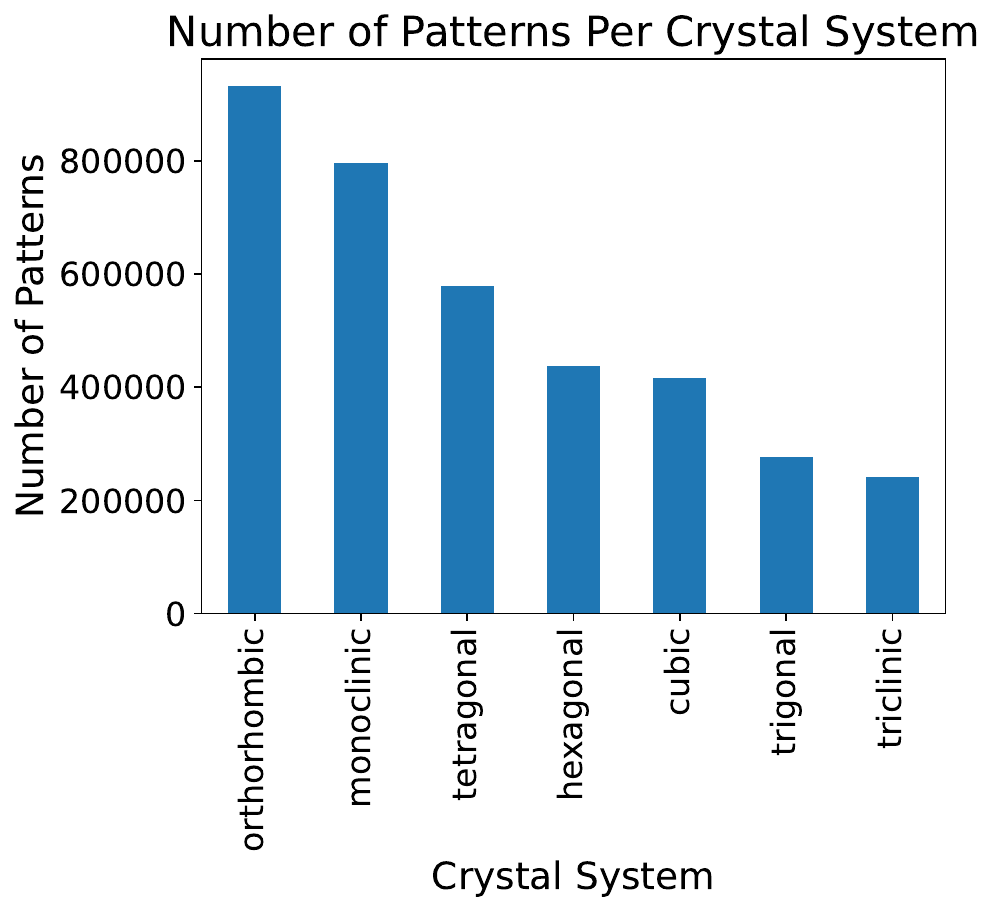}
    \caption{Number of simulated electron diffraction patterns per crystal system.}
    \label{fig_distribution} %This is for you to refer to the figure in maintext text
\end{figure*}

\begin{figure*}[ht]
    \centering
    \includegraphics[width=18cm]{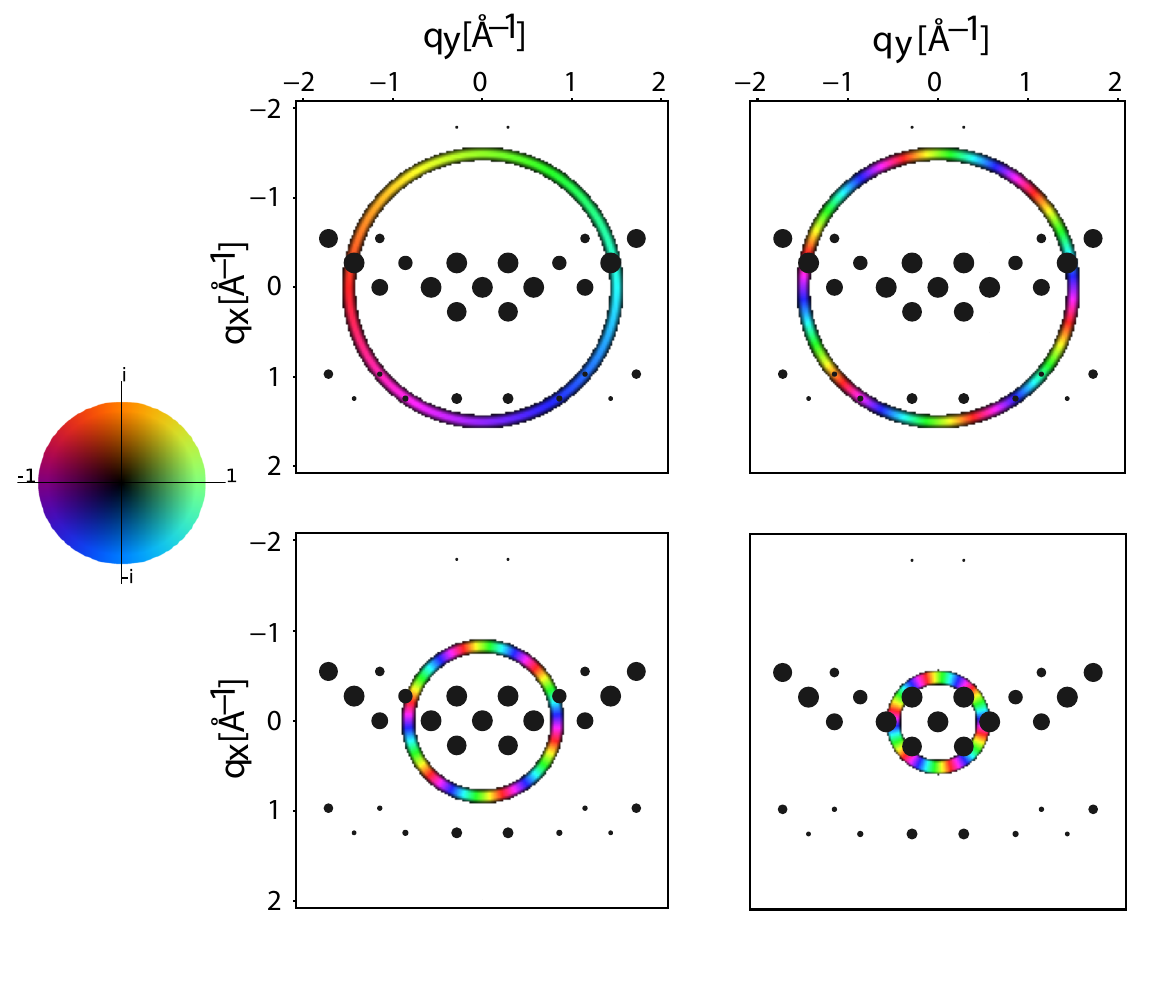}
    \caption{Illustration of radial vectorization of diffraction patterns. Every pattern is decomposed into an array of 41 radial vectors, indicated by the rings in each plot, and each contains from 0-12 angular modes, indicated by the number of full color oscillations each ring contains.}
    \label{fig_vectorization} %This is for you to refer to the figure in maintext text
\end{figure*}

\begin{figure*}[ht]
    \centering
    \includegraphics[width=18cm]{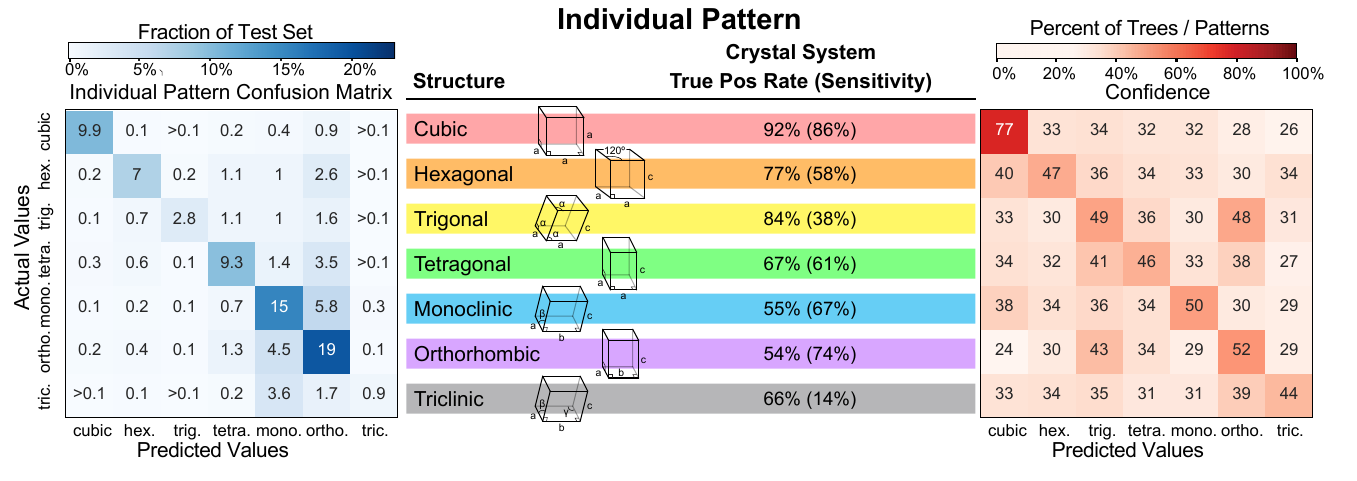}
    \caption{Summary results for the crystal system prediction model when triclinic is included in the training data. The confusion matrix on the left is not directly comparable to the accuracy confusion matrices in Figures 4 and 5 due to the different ratios of crystal systems in the full dataset due to the inclusion of triclinic. Therefore, all of the test set fractions will be lower than those found in Figures 4 and 5. In the crystal system column, the True Pos Rate is the percentage of correct predictions made of that crystal system.  For example, in the individual case 66\% of triclinic predictions are correct. The other 34\% are incorrectly called triclinic. The Sensitivity denotes the percentage of the total diffraction patterns within that crystal system correctly identified by the model. For example, in the individual case only 14\% of triclinic materials are labeled triclinic, while 86\% are labeled as some other crystal system. The extremely limited ability to identify triclinic, combined with the observation that including triclinic results in some false positives where non triclinic materials are called triclinic, resulted in triclinic’s exclusion from this model.}
    \label{triclinic} %This is for you to refer to the figure in maintext text
\end{figure*}

\begin{figure*}[ht]
    \centering
    \includegraphics[width=18cm]{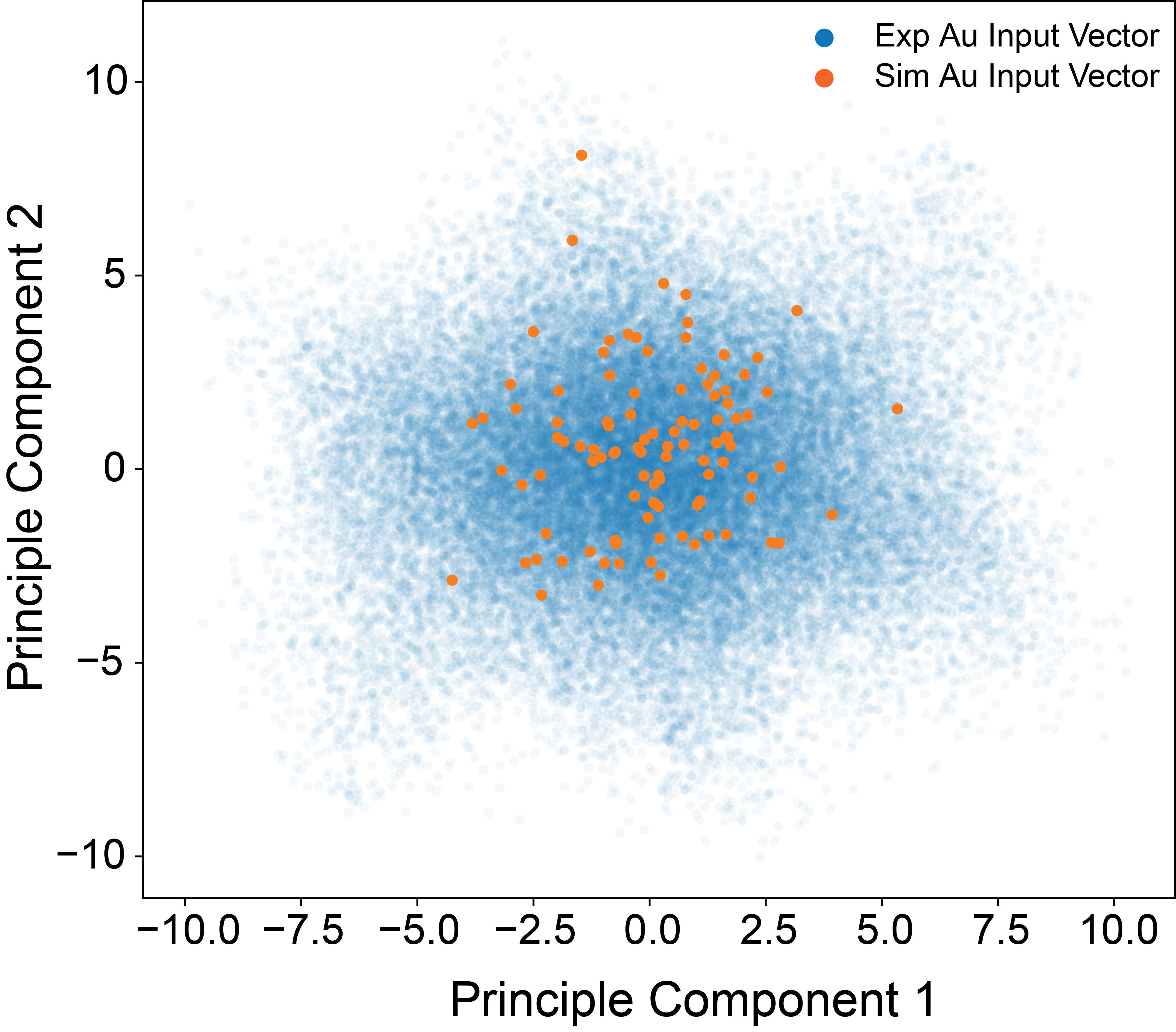}
    \caption{Principal component analysis (PCA) of the experimental electron diffraction data used to validate this model (blue) compared to the simulated electron diffraction patterns of FCC Au (orange) showing the first two principal components of the input vector. The experimental data is partially transparent, meaning darker blue regions are indicative of more data at those positions. }
    \label{pca_experiments} %This is for you to refer to the figure in maintext text
\end{figure*}

\begin{figure*}[ht]
    \centering
    \includegraphics[width=18cm]{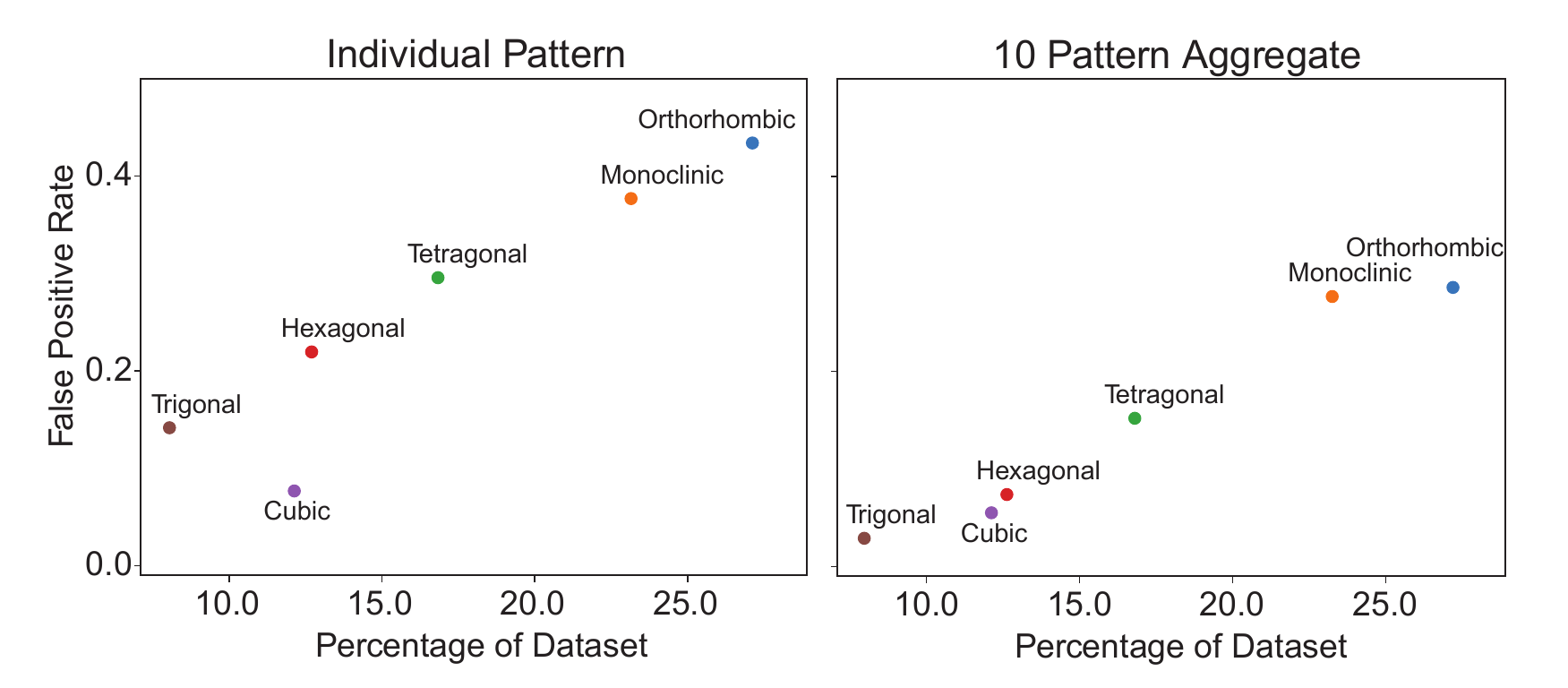}
    \caption{False positive rate vs percentage of the dataset represented by each crystal system.}
    \label{fig_false_positives} %This is for you to refer to the figure in maintext text
\end{figure*}

\begin{figure*}[ht]
    \centering
    \includegraphics[width=18cm]{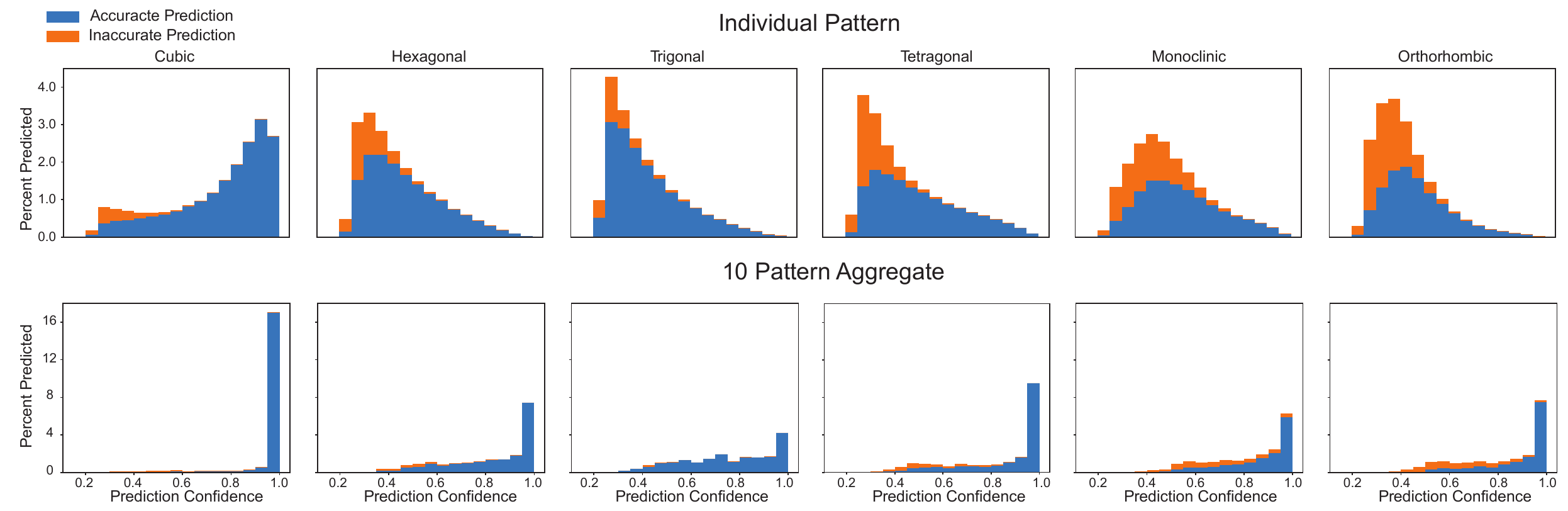}
    \caption{Spread of confidence levels for each entry in the confusion matrix for Figure 4c-d. The histograms show the frequency where a crystal system is predicted inaccurately (orange) stacked on top of correct predictions (blue).}
    \label{fig_confidences} %This is for you to refer to the figure in maintext text
\end{figure*}

\begin{figure*}[ht]
    \centering
    \includegraphics[width=18cm]{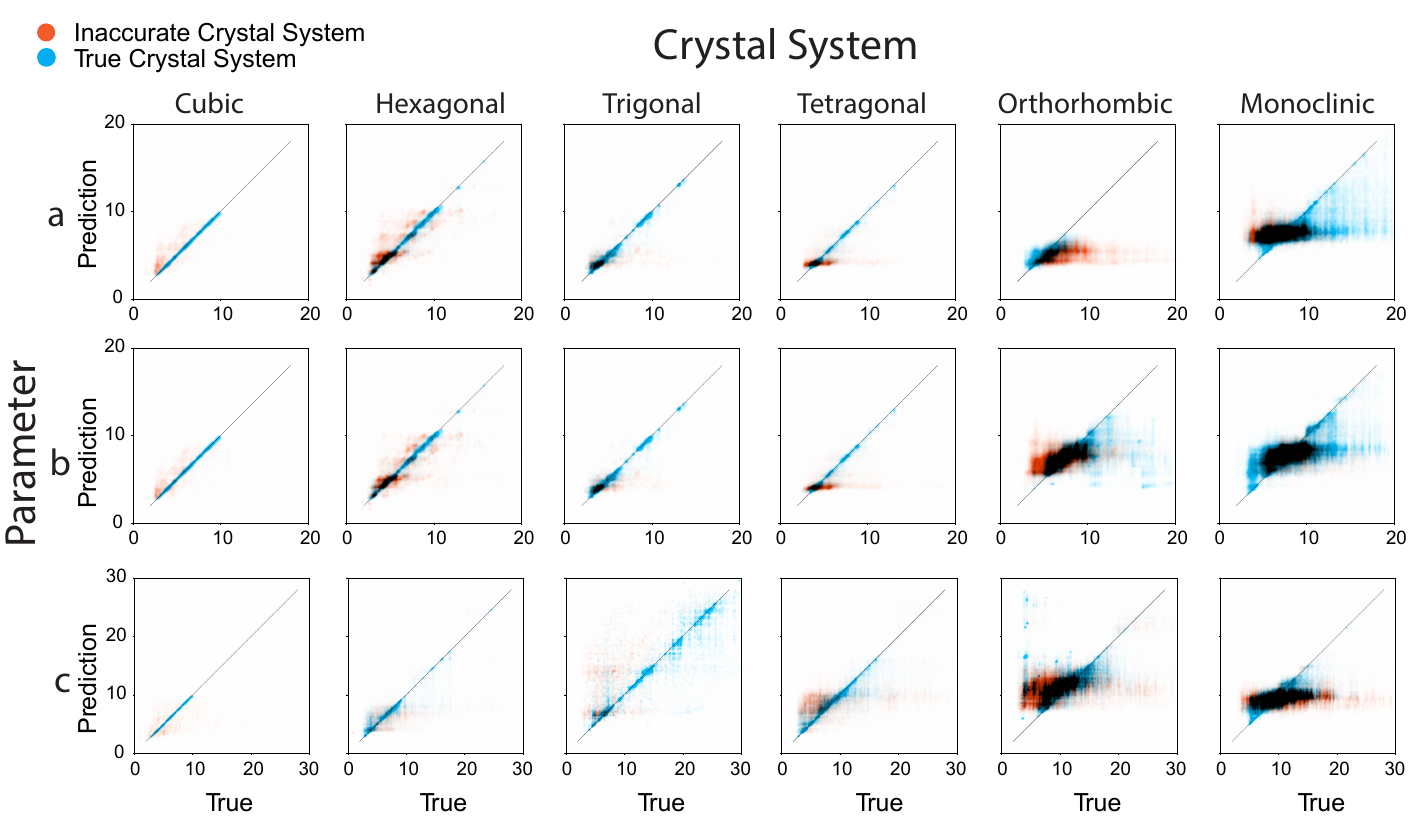}
    \caption{Prediction vs true lattice parameter prediction maps for individual pattern predictions. The blue regions show the accuracy when the crystal system has been predicted correctly while the orange regions show the accuracy when the crystal system has been incorrectly assigned to that class.}
    \label{fig_indvidual_lattice_r2} %This is for you to refer to the figure in maintext text
\end{figure*}

\begin{figure*}[ht]
    \centering
    \includegraphics[width=18cm]{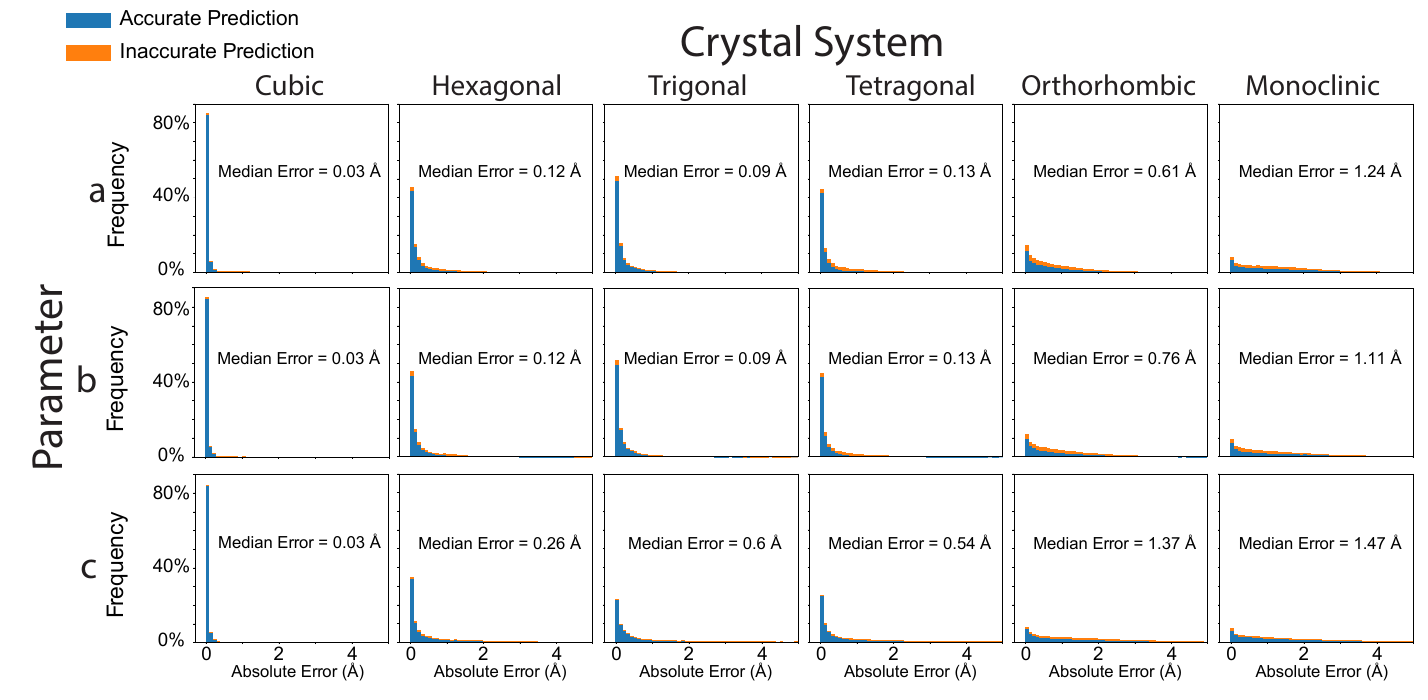}
    \caption{Error histograms for lattice parameter prediction from individual patterns. The histograms show the frequency where a crystal system is predicted inaccurately (orange) stacked on top of correct predictions (blue).}
    \label{fig_individual_lattice_MAE} %This is for you to refer to the figure in maintext text
\end{figure*}

\begin{figure*}[ht]
    \centering
    \includegraphics[width=18cm]{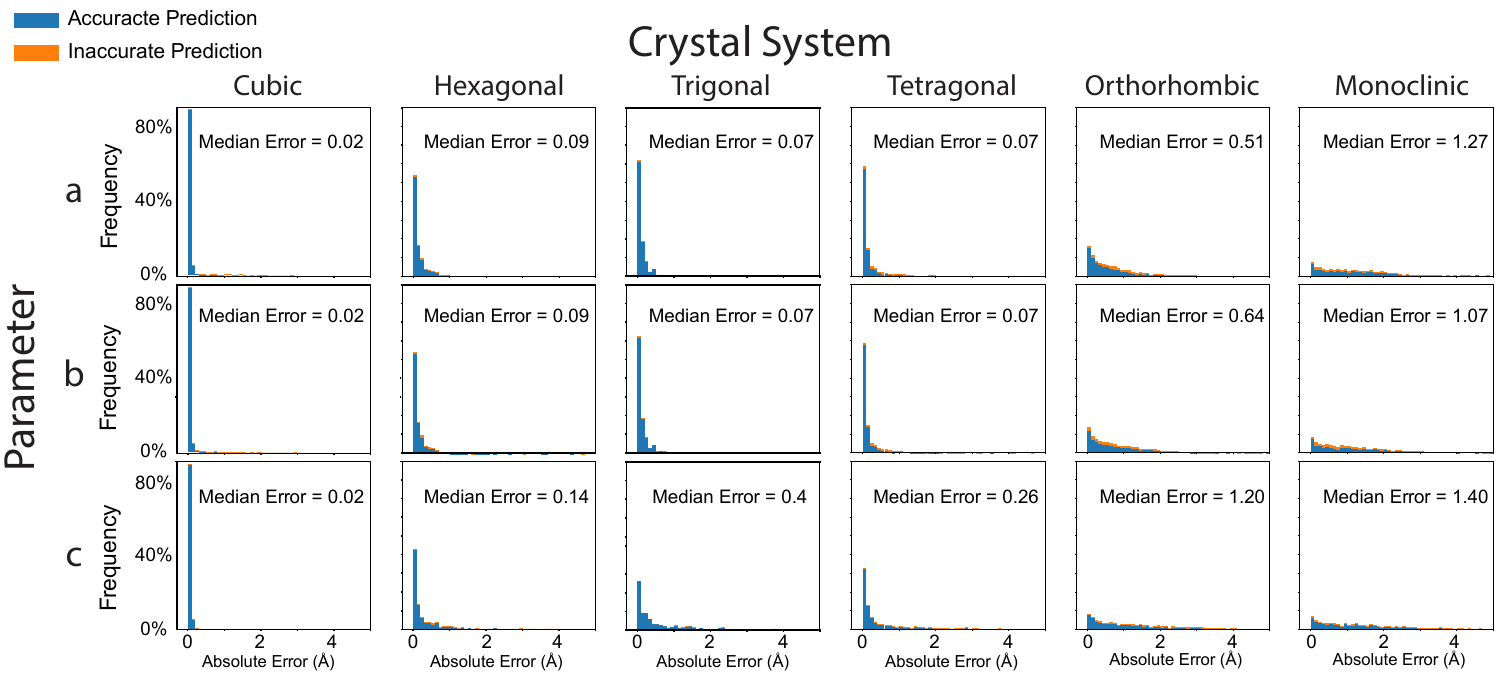}
    \caption{Error histograms for lattice parameter prediction from 10 aggregated patterns. The histograms show the frequency where a crystal system is predicted inaccurately (orange) stacked on top of correct predictions (blue).}
    \label{fig_aggregate_lattice_MAE} %This is for you to refer to the figure in maintext text
\end{figure*}

\begin{figure*}[ht]
    \centering
    \includegraphics[width=18cm]{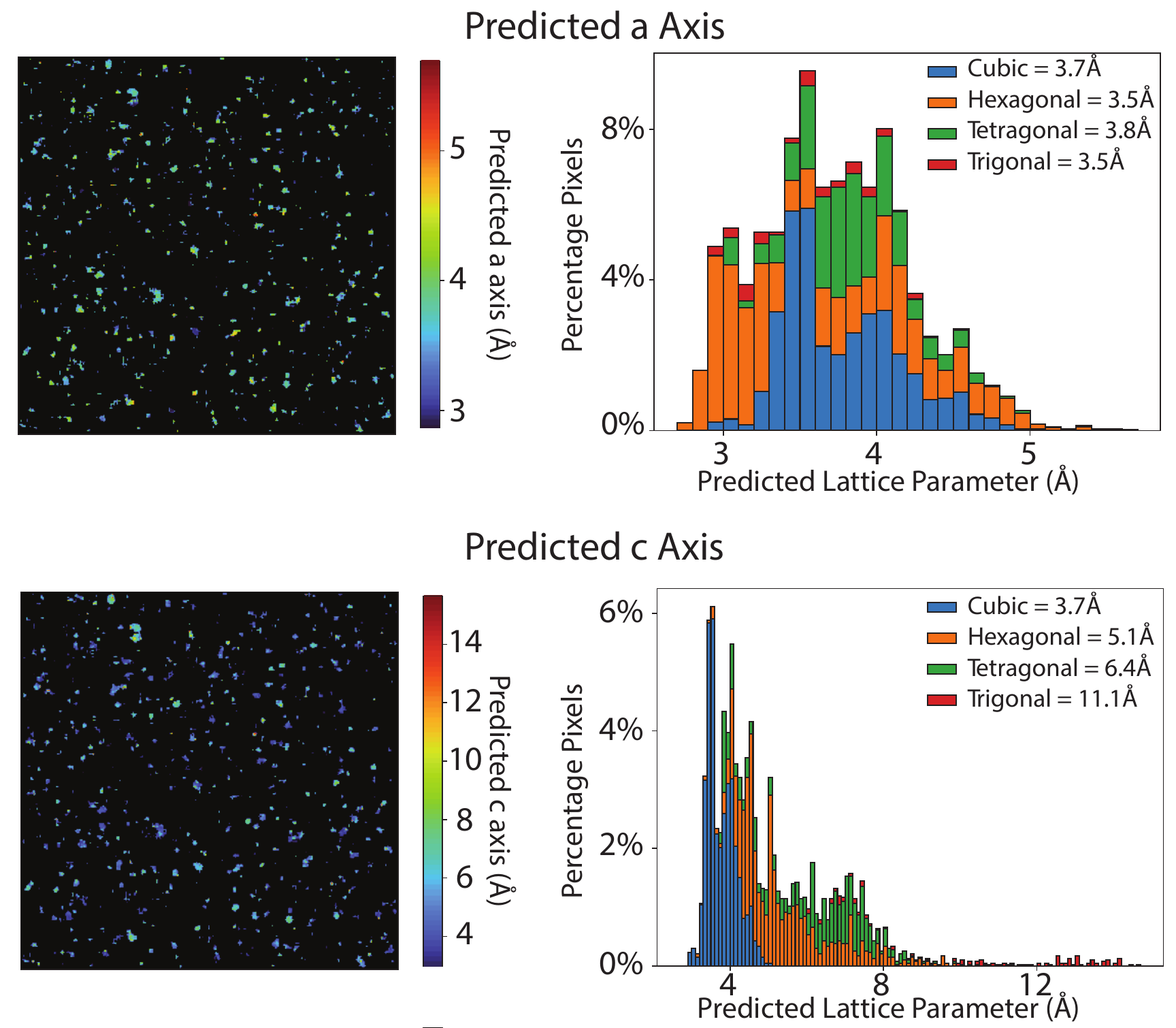}
    \caption{Lattice parameter predictions when the predicted crystal system lattice parameter model is used. The a and b axes are the same for all predicted crystal systems, therefore, only the predictions of the a and c axes are shown.}
    \label{fig_exp_lattice_full_pipeline} %This is for you to refer to the figure in maintext text
\end{figure*}